\documentclass[reprint,aps,prd]{revtex4-2}

\usepackage{ajtex-revtex}

\begin{document} 

\title{Fractonic solids}

\author{Akash Jain}\email{ajain@uva.nl}

\affiliation{Institute for Theoretical Physics, University of Amsterdam, 1090
  GL Amsterdam, The Netherlands}
\affiliation{Dutch Institute for Emergent Phenomena, 1090 GL Amsterdam, The Netherlands}
\affiliation{Institute for Advanced Study, University of Amsterdam, 1012 GC Amsterdam, The Netherlands}

\date{\today}

\begin{abstract}
Fractons are exotic quasiparticles whose mobility in space is restricted by symmetries. In potential real-world realisations, fractons are likely lodged to a physical material rather than absolute space. Motivated by this, we propose and explore a new symmetry principle that restricts the motion of fractons relative to a physical solid. Unlike models with restricted mobility in absolute space, these fractonic solids admit gauge-invariant momentum density, are compatible with boost symmetry, and can consistently be coupled to gravity. We also propose a holographic model for fractonic solids.
\end{abstract}

\maketitle

\tableofcontents

\section{Introduction}

Given a physical model with local degrees of freedom and local interactions, we typically expect its low-energy course-grained description to be reliably captured by a local effective field theory. However, in recent years, various lattice models of interest in condensed matter and high energy physics have emerged~\cite{Chamon:2004lew, 2011AnPhy.326..839B, Haah:2011drr, Vijay:2015mka}, whose low-energy descriptions feature exotic characteristics atypical of a well-posed effective field theory. The most striking characteristic of these models are local quasi-particle excitations, called \emph{fractons}, that cannot move freely in space despite translational invariance; see reviews~\cite{Nandkishore:2018sel, Pretko:2020cko}. 

The bizarre nature of fractons in many of these models can be ascribed to emergent \emph{subsystem symmetries} that act independently on different parts of the system~\cite{Vijay:2016phm, You:2018oai, Seiberg:2020bhn, Seiberg:2020wsg}. 
It is helpful to model these in terms of a series of \emph{multipole symmetries}~\cite{Pretko:2016kxt, Gromov:2018nbv}. Consider, e.g., a complex scalar field $\Psi$ invariant under following U(1) transformations
\begin{align}
    \Psi 
    \to {\rm e}^{i\Lambda_{\text{ss}}(\vec x)}\,\Psi
    = {\rm e}^{i\Lambda + i\Lambda_i x^i + \frac{i}{2}\Lambda_{ij}x^ix^j\ldots}\,\Psi.
    \label{eq:dipole-phase-orig}
\end{align}
The \emph{monopole symmetry} $\Lambda$ imposes charge conservation; the \emph{dipole symmetry} $\Lambda_i$ restricts the motion of free charges and only allows them to move in neutral bound states with fixed dipole moment; and similarly for higher moments. The associated symmetry algebra is characterised by the non-trivial commutation relations~\cite{Gromov:2018nbv}
\begin{equation}
    [\rmP_i,\rmQ^j] = i\delta_i^j\rmQ, \quad 
    [\rmP_i,\rmQ^{jk}] = 2i\delta_i^{(j}\rmQ^{k)}, \quad \ldots,
    \label{eq:symm-algebra-spatial}
\end{equation}
between the momenta $\rmP_i$ and the multipole generators $\rmQ,\rmQ_{i},\rmQ_{ij},\ldots$. This implies, for instance, that translations do not commute with dipole transformations when acting on charged states. 

A byproduct of restricted spatial mobility is that multipole symmetries do not play well with spacetime symmetries. They are outright incompatible with boost transformations and thus cannot be realised in relativistic or Galilean systems.
In particular, multipole-invariant field theories cannot consistently be coupled to general relativistic gravitational spacetimes~\cite{Slagle:2018kqf, Jain:2021ibh}. They can be made compatible with spacetime translations, but require exotic boost-agnostic spacetime geometries to couple to the associated energy-momentum currents~\cite{Pena-Benitez:2021ipo, Jain:2021ibh, Bidussi:2021nmp}. Moreover, momentum is not invariant under multipole transformations in these theories and leads to peculiarities such as no low-energy phases with spontaneously unbroken multipole symmetries~\cite{Glorioso:2023chm, Jensen:2022ibn} and lack stable hydrodynamic states with nonzero velocity~\cite{Gromov:2020yoc, Osborne:2021mej, Grosvenor:2021rrt, Jain:2023nbf, Armas:2023ouk, Jain:2024kri}.

\begin{figure}
    \centering
    \includegraphics[width=100px]{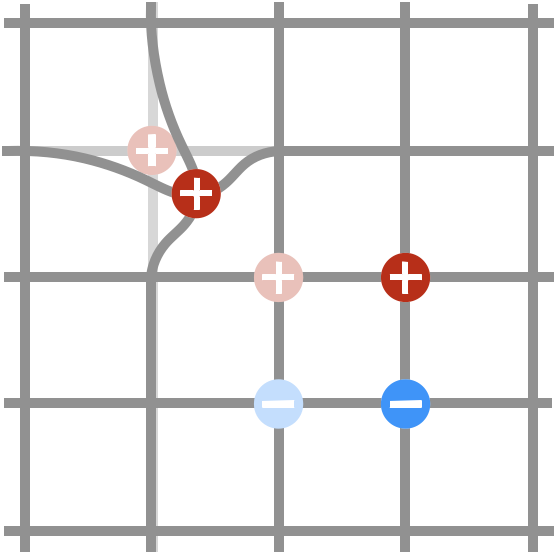}
    \caption{Crystal-dipole symmetry. Isolated free charges cannot move on the crystal, but can move in space when the crystal is distorted. Neutral bound states can move freely on the crystal while keeping their crystal-dipole moment fixed.}
    \label{fig:crystal-dipole}
\end{figure}

While fractons remain largely a theoretical curiosity, various condensed matter systems are expected to realise fractonic excitations at low energies. Examples include topological defects in crystals and superfluids~\cite{Pretko:2018qru, Nguyen:2020yve, Gromov:2017vir, Caddeo:2022ibe, Doshi:2020jso}, majorana islands \cite{You:2018bmf}, plaquette paramagnets~\cite{You:2019cvs}, hole-doped antiferromagnets~\cite{Sous:2019jtt}, and tilted optical lattices~\cite{Guardado-Sanchez:2019bjm}. In such physical realisations, one expects that fractons are embedded on a physical material but they may move in absolute space when the material itself is strained or distorted. See \cref{fig:crystal-dipole}. Accounting for the dynamics of the underlying material may also help ease tensions between multipole and spacetime symmetries. As a first step in this direction, the goal of this letter is to investigate effective field theories for fractons whose motion is restricted with respect to a dynamical crystal. 

\section{Crystal-multipole symmetries}

We cannot naively append elastic crystal dynamics to the multipole symmetries \eqref{eq:symm-algebra-spatial}. This would correspond to an unnatural scenario where fractons stay lodged in absolute space while the crystal is distorted. We need to instead modify the multipole symmetry itself to make it aware of the crystalline structure. A crystal in $d$ spatial dimensions is described by a collection of \emph{crystal fields} $\phi^{I=1,\ldots,d}$ representing the Eulerian coordinates of the underlying lattice sites as a function of time and space $(x^\mu) = (t,x^i)$~\cite{Leutwyler:1996er, Nicolis:2015sra, Armas:2019sbe, Armas:2020bmo, Armas:2021vku, Armas:2022vpf}. To impose mobility restrictions on the crystal, we adapt the multipole transformations in \cref{eq:dipole-phase-orig} using $\phi^I$ as
\begin{equation}
    \Psi 
    \to {\rm e}^{i\Lambda_{\text{ss}}(\phi^I(x))}\,\Psi
    = {\rm e}^{i\Lambda + i\Lambda_I\phi^I(x) + \frac{i}{2}\Lambda_{IJ}\phi^I(x)\phi^J(x) + \ldots}\,\Psi,
    \label{eq:dipole-phase-crys}
\end{equation}
which reduces to \cref{eq:dipole-phase-orig} when the crystal is fixed to an unstrained equilibrium state $\phi^I = \delta^I_i x^i$. However, beyond this simple state, the \emph{crystal-multipole} symmetries \eqref{eq:dipole-phase-crys} depend non-linearly on the crystal configuration and thus qualitatively differ from the spatial-multipole symmetries \eqref{eq:dipole-phase-orig}. The spatial-momenta $\rmP_i$ in the symmetry algebra \eqref{eq:symm-algebra-spatial} are appropriately replaced by the crystal momenta $\rmP_I$ that generate crystal translations $\phi^I\to \phi^I+a^I$, i.e.
\begin{equation}
    [\rmP_I,\rmQ^J] = i\delta_I^J\rmQ, \quad 
    [\rmP_I,\rmQ^{JK}] = 2i\delta_I^{(J}\rmQ^{K)}, \quad \ldots.
    \label{eq:symm-algebra-crystal}
\end{equation}
See more details regarding the symmetry algebra in \cref{app:crystal-dipole,app:crystal-multipole}.
Since the crystal-multipole transformations \eqref{eq:dipole-phase-crys} only talk to the spacetime coordinates via the crystal fields, they get along better with spacetime symmetries. In particular, field theories with crystal-multipole symmetries are compatible with boosts, can be realised in relativistic or Galilean systems, can be coupled to dynamical gravity, admit translationally-invariant phases with spontaneously unbroken crystal-multipole symmetries, and allow stable states with nonzero velocity. We will investigate all these aspects in detail in this letter, specialising to relativistic systems with crystal-dipole symmetry.

\section{Field theory with crystal-dipole symmetry}

A crystal naturally defines the spacelike frame fields $e^I_\mu = \partial_\mu\phi^I$ and the timelike frame velocity $v^\mu$ s.t. $v^\mu e^I_\mu = 0$, $v^\mu v_\mu= -1$. Raising/lowering of spacetime and crystal indices is performed using the relativistic spacetime metric $g_{\mu\nu}$ and the crystal metric $h^{IJ}=e^{I\mu}e^J_{\mu}$ respectively; see~\cite{Fukuma:2011pr, Armas:2020bmo}. Using this dynamical segregation between time and space, and drawing motivation from spatial-dipole-invariant field theories in~\cite{Pretko:2018jbi}, we can engineer a crystal-dipole-invariant Lagrangian for a relativistic complex scalar field $\Psi$, e.g.
\begin{equation}
    {\cal L}_\Psi 
    = |\Df_v \Psi|^2
    - \lambda\, |\rmD_{IJ}(\Psi,\Psi)|^2
    - V(|\Psi|^2),
    \label{eq:complex-scalar-L}
\end{equation}
for some parameter $\lambda$ and potential $V(|\Psi|^2)$. The derivative operators are defined as
\begin{align}
    \Df_{IJ}(\Psi,\Psi) 
    &= \Psi\Df_{(I}\Df_{J)} \Psi - \Df_I\Psi\Df_J\Psi,\nn\\
    \Df_v 
    &= v^\mu \nabla_\mu, \qquad 
    \Df_I = e^{\mu}_I\nabla_\mu, 
    \label{eq:derivative-operators-flat}
\end{align}
and $\nabla_\mu$ is the spacetime covariant derivative. One may explicitly check that the theory is invariant under global symmetries: crystal-translations, crystal-rotations, monopole and crystal-dipole transformations, as well as spacetime Poincar\'e transformations. In particular, a quadratic spatial derivative term $|\Df_I\Psi|^2$ is forbidden by crystal-dipole symmetry. For a fixed homogeneous crystal, $\phi^I = \delta^I_ix^i$, this theory reduces to the model of~\cite{Pretko:2018jbi} with spatial-dipole symmetry and explicitly broken Lorentz-invariance.

To account for crystal dynamics, the Lagrangian also contains an elastic part
\begin{align}
    \cL_{\text{el}} 
    = - \half \lb B-{\txp\frac{2}{d}}G \rb (\kappa^I_{~I})^2
    - G\, (\kappa_{IJ})^2,
    \label{eq:L-el}
\end{align}
where the crystal strain tensor $\kappa_{IJ} = \half(h_{IJ}-\delta_{IJ})$ measures distortions away from the equilibrium configuration, and $B$, $G$ are the elastic moduli.

We can extremise the effective action \smash{$S=\int_{d+1}\cL$} with respect to the dynamical fields $\Psi$ and  $\phi^I$ to obtain the respective equations of motion. Here $\int_{d+1}$ is shorthand for the integration measure $\int\df t\,\df^{d}\vec x\,\sqrt{-\det g}$. The equations of motion have been derived in \cref{app:complex-scalar}.

\section{Gauge fields and conserved currents}

We can gauge the monopole and crystal-dipole symmetries discussed above by introducing gauge fields $A_\mu$, $A_{I\mu}$ that transform as
\begin{align}
    A_\mu &\to A_\mu + \dow_\mu \Lambda  - a^I A_{I\mu}, \quad
    A_{I\mu}\to A_{I\mu} + \dow_\mu \Lambda_I.
    \label{eq:gauge-trans}
\end{align}
The shift of $A_\mu$ under crystal translations $a^I$ is required by the crystal-dipole symmetry algebra in \cref{eq:symm-algebra-crystal}. More details on the gauging procedure can be found in \cref{app:crystal-dipole}. The gauge-covariant derivative of $\Psi$ is defined as $\Df_\mu \Psi = \dow_\mu \Psi - i\lb A_\mu + \phi^I A_{I\mu} \rb \Psi$, extending the derivative operators in \cref{eq:derivative-operators-flat} to 
\begin{align}
    \Df_{IJ}(\Psi,\Psi) 
    &= \Psi\Df_{(I}\Df_{J)} \Psi - \Df_I\Psi\Df_J\Psi 
    - ie^{\mu}_{(I} A_{J)\mu} \Psi^2, \nn\\
    \Df_v 
    &= v^\mu \Df_\mu, \qquad 
    \Df_I = e^{\mu}_I \Df_\mu.
\end{align}
With these gauged definitions of the derivative operators, the complex scalar field theory in \cref{eq:complex-scalar-L} becomes invariant under local monopole and crystal-dipole transformations.

Varying the effective action $S=\int_{d+1}\cL$ of the theory with respect to the background metric and gauge fields, we can read off the respective conserved currents: symmetric energy-momentum tensor $T^{\mu\nu}$, charge monopole current $J^\mu$, and crystal-dipole current $J^{\mu I}$, i.e.
\begin{equation}
    \delta S 
    = \int_{d+1} \lb
    \half T^{\mu\nu}\delta g_{\mu\nu} 
    + J^\mu \delta A_\mu
    + \left(J^{\mu I}+J^\mu \phi^I\right) \delta A_{I\mu} \rb.
    \label{eq:action-variation}
\end{equation} 
We have isolated the ``free'' contribution $J^\mu \phi^I$ from the ``internal'' dipole current $J^{\mu I}$ to make it invariant under crystal-translations. The conservation equations are given as
\begin{gather}
    \nabla_\mu T^{\mu\nu}
    = \tilde F^{\nu\rho} J_\rho
    + F_I^{\nu\rho} J^I_\rho,\nn\\
    \nabla_\mu J^\mu = 0,\qquad
    \nabla_\mu J^{\mu I} = -e^I_\nu J^\nu,
    \label{eq:cons-free}
\end{gather}
where we have defined the monopole and crystal-dipole gauge field strengths $F_{\mu\nu} = 2\dow_{[\mu}A_{\nu]}$, $F_{I\mu\nu} = 2\dow_{[\mu}A_{I\nu]}$, together with the combination $\tilde F_{\mu\nu} = F_{\mu\nu} + \phi^I F_{I\mu\nu}$ that is invariant under crystal-translations. The third conservation equation is the crystalline incarnation of the dipole Ward identity, which relates the divergence of the internal crystal-dipole current to the monopole flux along the crystal. The total crystal-dipole moment
\begin{align}
    \int_d J^0\phi^I + J^{0I},
\end{align}
is conserved, together with the total energy-momentum and charge, where $\int_d$ is $\int\df^d\vec x\,\sqrt{-g} $. Note that the total spatial-dipole moment $\int_d J^0 x^i + J^{0I}\delta_I^i$ is not conserved. Therefore, monopole charges can move in absolute space as long as the generated dipole moment is compensated by the elastic contribution $\int_d J^0\delta_I^i\delta\phi^I$.
For the complex scalar field theory in \eqref{eq:complex-scalar-L}, $J^\mu$ and $J^{\mu I}$ are given as
\begin{align}\label{eq:scalar-field-currents}
    J^\mu 
    &= -iv^\mu\Psi\Df_v\Psi^*
    - i\lambda e^\mu_{I}\, 
    \dow_{\nu}\Big(e^\nu_J\Df^{\dagger IJ}(\Psi,\Psi) \Psi^2 \Big)
    + \text{c.c.}
    , \nn\\
    J^{\mu I}
    &= i\lambda e^\mu_{J} \Df^{\dagger IJ}(\Psi,\Psi) \Psi^2
        + \text{c.c.},
\end{align}
where ``c.c.'' denotes the complex conjugate. Similarly, the energy-momentum tensor $T^{\mu\nu}$ is given as
\begin{align}\label{eq:em-tensor}
    T^{\mu\nu}
    &= |\Df_v \Psi|^2 v^\mu v^\nu
    + 2\lambda \rmD^{\dagger}_{IK}(\Psi,\Psi) \rmD_J^{~K}(\Psi,\Psi) e^{I\mu}e^{J\nu} \nn\\
    &~~
    - 2\lambda \rmD^{\dagger IJ}(\Psi,\Psi) \bigg( 2\Df_{v(I}(\Psi,\Psi)
    + \Psi \Df_v\Psi\, \dow_{I}
    \bigg) 
    v^{(\mu} e_{J}^{\nu)} \nn\\
    &~~
    + \frac12 \cL_\Psi g^{\mu\nu}
    + \text{c.c.}.
\end{align}
See \cref{app:complex-scalar} for more details. In addition, $T^{\mu\nu}$ also receives an elastic contribution arising from $\cL_{\text{el}}$ in \cref{eq:L-el}; see e.g.~\cite{Armas:2019sbe}.

\section{Crystal-dipole gauge theory}

We can give dynamics to the monopole and crystal-dipole gauge fields by introducing new kinetic terms in the Lagrangian, e.g.
\begin{align}\label{eq:gauge-theory}
    {\cal L}_{\text{gauge}}
    &= - \frac{1}{4g_0}\tilde F_{\mu\nu} \tilde F^{\mu\nu}
    - \frac{1}{4g_1} F_{I\mu\nu} F^{I\mu\nu},
\end{align}
where $g_0$, $g_1$ are constants.
More general terms may be included by projecting the field strengths along $v^\mu$ and $e_\mu^I$.
The respective Maxwell-type equations and Bianchi identities are given as
\begin{alignat}{2}\label{eq:maxwell}
    \nabla_\nu\tilde F^{\mu\nu}
    &= g_0 J^\mu, &
    \nabla_\nu F^{I\mu\nu} 
    &= g_1J^{I\mu}
    - \frac{g_1}{g_0} e^I_\nu \tilde F^{\mu\nu}, \nn\\
    \nabla_{[\mu}\tilde F_{\nu\rho]} 
    &= e^I_{[\mu}F_{I\nu\rho]}, \qquad &
    \nabla_{[\mu}F_{I\nu\rho]} 
    &= 0.
\end{alignat}
These may be understood as a relativistic extension of the symmetry-tensor gauge theory studied in~\cite{Pretko:2018jbi}, embedded on a dynamical crystal. To wit, the field strength components $\tilde F_{\mu\nu}$, $F_{[ij]t}$, $F_{[ijk]}$, where $F_{\mu\nu\rho} = e_I^\mu F_{I\nu\rho}$, are gapped due to the second and third equations in \cref{eq:maxwell}. 
Taking $d=3$, the remaining gapless components are contained in the tensor analogues of electric and magnetic fields
\begin{align}
    E_{ij}=F_{(ij)t} + \dow_{(i}\tilde F_{j)t}, \quad 
    B_i^{~j}=\half\epsilon^{jkl}(F_{ikl}-\dow_k\tilde F_{li}),    
\end{align}
satisfying $E_{[ij]},B_i^{~i}=0$,
which behave analogous to the symmetric-tensor gauge theory at the linearised level. However, the dynamics differs significantly at the non-linear level due to interactions with the crystal and depends on the explicit form of the Lagrangian \eqref{eq:gauge-theory}.

\section{Spontaneous symmetry breaking}

As a phase of matter, crystals or solids spontaneously break spatial translation symmetry, with $\phi^I$ identified as the respective Goldstone fields. Analogous to the supersolid phase that further spontaneously breaks the U(1) particle conservation symmetry~\cite{2012RvMP...84..759B}, we may consider \emph{multipole supersolids} with spontaneously broken crystal-multipole symmetries. Following~\cite{Stahl:2023prt, Jensen:2022ibn}, there are two distinct possibilities for systems with crystal-dipole symmetry: a \emph{p-wave} phase with $\langle\Psi\rangle=0$, $\langle i\Psi^\dagger\Df_I\Psi\rangle\neq 0$ where the crystal-dipole symmetry is spontaneously broken but the monopole remains unbroken, and an \emph{s-wave} phase with $\langle\Psi\rangle\neq 0$ where both are spontaneously broken.

The p-wave phase is characterised by a vector Goldstone field $\varphi_I$, transforming as $\varphi_I \to \varphi_I - \Lambda_I$. Defining the crystal-dipole superfluid velocity $\xi_{I\mu} = \dow_\mu\varphi_I + A_{I\mu}$, we may write down a simple effective Lagrangian for the p-wave phase as
\begin{align}\label{eq:pwave-Lagrangian}
    {\cal L}_{\text{p-wave}}
    &= \half\chi_{\rm d}(v^\mu\xi_{I\mu})^2
    + V({\txp\xi_{I}^{~J}}),
\end{align}
where the potential $V$ depends on $\xi_{IJ} = \xi_{I\mu} e^{J\mu}$.

In contrast, the s-wave phase is characterised by a scalar Goldstone field $\varphi$, transforming as
$\varphi \to \varphi - \Lambda - \Lambda_I\phi^I$. The second term in the transformation is necessitated by the symmetry algebra \eqref{eq:symm-algebra-crystal}. As a consequence, the superfluid velocity $\xi_\mu = \dow_\mu\varphi + A_\mu + \phi^I A_{I\mu}$ is invariant under the monopole symmetry but not the crystal-dipole symmetry. In fact, $\xi_I\equiv e_I^\mu \xi_\mu$ transforms exactly like the vector Goldstone $\varphi_I$, justifying that the crystal-dipole symmetry is automatically spontaneously broken, and we may use it to define $\xi_{I\mu} = \dow_\mu\xi_I +  A_{I\mu}$. Using these ingredients, we may construct the effective Lagrangian, e.g.
\begin{align}
    {\cal L}_{\text{s-wave}}
    &= \half\chi\,(v^\mu\xi_\mu)^2 
    + \half\chi_{\rm d}(v^\mu\xi_{I\mu})^2
    + V({\txp\xi_{I}^{~J}}),
\end{align}
which is similar in form to \cref{eq:pwave-Lagrangian}, but with a new kinetic term $\sim(v^\mu\xi_\mu)^2$ for the scalar Goldstone.

\section{Crystal-dipole hydrodynamics}

Imposing mobility restrictions on a dynamical crystal instead of absolute space has qualitative impact on the low-energy thermal phases. The dynamics of a system at finite temperature is described by the framework of hydrodynamics~\footnote{One may also approach hydrodynamics using Schwinger-Keldysh effective actions~\cite{Grozdanov:2013dba, Harder:2015nxa, Crossley:2015evo, Haehl:2015uoc, Haehl:2018lcu, Jensen:2017kzi, Glorioso:2018wxw}, useful for including stochastic fluctuations in hydrodynamic models.}. Therein, one specifies the constitutive relations for the conserved currents $T^{\mu\nu}$, $J^\mu$, $J^{\mu I}$ in terms of the hydrodynamic variables: fluid velocity $u^\mu$ (s.t. $u^\mu u_\mu=-1$), temperature $T$, chemical potential $\mu$, crystal-dipole chemical potential $\mu_I$, and the crystal fields $\phi^I$. To wit, consider
\begin{align}\label{eq:consti}
    T^{\mu\nu} 
    &= (\epsilon+p)u^\mu u^\nu + p\,g^{\mu\nu}
    - r^{IJ}e^{\mu}_I e^{\nu}_J \nn\\
    &\qquad 
    - 2\eta \Delta^{\mu\rho}\Delta^{\nu\sigma} \nabla_{(\rho}u_{\sigma)}
    - \lb\zeta - {\txp\frac{2}{d}}\eta\rb \Delta^{\mu\nu}\nabla_\lambda u^\lambda
    , \nn\\
    J^\mu
    &= n\,u^\mu
    - \sigma \Delta^{\mu\nu}\lb T\dow_\nu \frac{\mu}{T}
    - \mu_I e^I_\nu - \tilde F_{\nu\rho}u^\rho \rb, \nn\\
    J^{\mu I}
    &= n^I u^\mu - \sigma_{\rm d} \Delta^{\mu\nu} h^{IJ}
    \lb T\dow_\nu \frac{\mu_J}{T} - F_{J\nu\rho} u^\rho \rb.
\end{align}
The energy density $\epsilon$, pressure $p$, entropy density $s$, charge density $n$, crystal-dipole density $n^I$, and elastic stress $r_{IJ}$ obey the thermodynamic relations
\begin{align}
    \df p &= s\,\df T + n\,\df\mu + n^I\df\mu_I + \half r_{IJ} \df h^{IJ}, \nn\\
    \epsilon &= Ts + \mu n + \mu_I n^I - p.
\end{align}
We have also introduced the dissipative viscosities $\eta$, $\zeta$ and conductivities $\sigma$, $\sigma_{\rm d}$. The conservation equations \eqref{eq:cons-free} govern the dynamics of $u^\mu$, $T$, $\mu$, $\mu_I$, while that of $\phi^I$ is governed by the Josephson equation
\begin{align}\label{eq:KI}
    K_I 
    &\equiv - \nabla_\mu\!\lb r_{IJ} e^{J\mu} \rb 
    + n \mu_I
    - \sigma_\phi u^\mu e_{I\mu}
    = 0,
\end{align}
with the crystal diffusion parameter $\sigma_\phi$.
The form of hydrodynamic equations is dictated by the local second law of thermodynamics; see \cref{app:second-law} for more details. However, we have only included a selection of transport coefficients here for simplicity.

\newcommand{\centext}[1]{$$\begin{aligned}
\text{#1}
\end{aligned}$$}

\begin{table*}
    \centering
    \begin{tabular}{p{3.5cm}||p{4.5cm}|p{4.5cm}|p{4.5cm}|}
        & \vspace{-1em} \centext{Ordinary phase}
        & \vspace{-1em} \centext{s-wave phase}
        & \vspace{-1em} \centext{p-wave phase} \\[-1.5em]
        \hline\hline
        \centext{Charged fluids}
        & $$\begin{aligned}
            \omega_{\text{sound}}^\pm &\sim \pm k -i k^2 \\
            \omega^\perp_{\text{shear}} &\sim -i k^2 \\
            \omega_{\text{charge}} &\sim -i k^2
        \end{aligned}$$
        & $$\begin{aligned}
            \omega_{\text{sound}}^\pm &\sim \pm k -i k^2 \\
            \omega^\perp_{\text{shear}} &\sim -i k^2 \\
            \omega_{\text{2nd-sound}}^\pm &\sim \pm k -i k^2
        \end{aligned}$$
        & \centext{n/a} \\
        \hline 
        \centext{Charged solids}
        & $$\begin{aligned}
            \omega_{\text{sound}}^\pm &\sim \pm k -i k^2 \\
            \omega^{\perp\pm}_{\text{phonon}} &\sim \pm k -i k^2 \\
            \omega_{\text{crystal}} &\sim -i k^2 \\
            \omega_{\text{charge}} &\sim -i k^2
        \end{aligned}$$
        & $$\begin{aligned}
            \omega_{\text{sound}}^\pm &\sim \pm k - ik^2 \\
            \omega^{\perp\pm}_{\text{phonon}} &\sim \pm k -i k^2 \\
            \omega_{\text{2nd-sound}}^\pm &\sim \pm k -i k^2 \\
            \omega_{\text{crystal}} &\sim -i k^2
        \end{aligned}$$
        & \centext{n/a} \\
        \hline
        $$\begin{gathered}
        \text{Spatial-dipole fluids}\\
        \text{(no boost symmetry)}
        \end{gathered}$$
        & \centext{Phase does not exist}
        & $$\begin{aligned}
            \omega_{\text{sound}}^\pm &\sim \pm k - ik^2 \\
            \omega^\perp_{\text{shear}} &\sim -i k^2 \\
            {\color{red}\omega_{\text{2nd-sound}}^\pm} &~
            {\color{red}\sim \pm k^2 - ik^4} \hspace{-2em}
        \end{aligned}$$
        & $$\begin{aligned}
            {\color{red}\omega_{\text{sound}}^\pm} &~
            {\color{red}\sim \pm k^2 - ik^2}\hspace{-2em} \\
            {\color{red}\omega^\perp_{\text{shear}}} &~
            {\color{red}\sim -ik^4} \\
            \omega_{\text{charge}} &\sim -i k^2
        \end{aligned}$$ \\
        \hline
        \centext{Crystal-dipole solids}
        & $$\begin{aligned}
            \omega_{\text{sound}}^\pm &\sim \pm k -i k^2 \\
            \omega^{\perp\pm}_{\text{phonon}} &\sim \pm k -i k^2 \\
            \omega_{\text{crystal}} &\sim -i k^2 \\
            {\color{red}\omega^\perp_{\text{charge}}} &~
            {\color{red}\sim -ik^4}
        \end{aligned}$$
        & $$\begin{aligned}
            \omega_{\text{sound}}^\pm &\sim \pm k - ik^2 \\
            \omega^{\perp\pm}_{\text{phonon}} &\sim \pm k -i k^2 \\
            \omega_{\text{crystal}} &\sim -i k^2 \\
            {\color{red}\omega_{\text{2nd-sound}}^\pm} &~
            {\color{red}\sim \pm k^2 - ik^4} \hspace{-2em}
        \end{aligned}$$
        & $$\begin{aligned}
            \omega_{\text{sound}}^\pm &\sim \pm k  - ik^2 \\
            \omega^{\perp\pm}_{\text{phonon}} &\sim \pm k -i k^2 \\
            \omega_{\text{crystal}} &\sim -i k^2 \\
            {\color{red}\omega^\perp_{\text{p-wave}}} &~
            {\color{red}\sim -ik^2} \\
            {\color{red}\omega_{\text{2nd-sound}}^\pm} &~
            {\color{red}\sim \pm k^2 - ik^4} \hspace{-2em} \\
            &
            {\color{red}~\small{\text{or}}~ {-ik^2}} \\
        \end{aligned}$$ \\
        \hline
    \end{tabular}
    \caption{Comparison of the mode spectrum of charged fluids, charged solids, spatial-dipole-conserving fluids (without boost symmetry), and crystal-dipole-conserving fluids. We have highlighted the fractonic modes (i.e. magnon-like propagating modes $\omega\sim\pm k^2$ and subdiffusive modes $\omega\sim -ik^4$) in red. The transverse modes denoted with ``$\perp$'' appear in $d-1$ copies for each of the transverse directions.}
    \label{tab:modes}
\end{table*}

Let us look at the spectrum of linearised hydrodynamic fluctuations around the equilibrium solution: $u^\mu = \delta^\mu_t$, $T=T_0$, $\mu=\mu_0$, $\mu_I=0$, and $\phi^I = \delta^I_i x^i$, in the absence of background sources. For an ordinary crystal, the spectrum contains longitudinal sound modes $\omega\sim\pm k$, transverse phonon modes $\omega\sim\pm k$, and two coupled charge and crystal diffusion modes $\omega\sim -ik^2$.
However, due to crystal-dipole symmetry, the charge diffusion mode decouples and becomes subdiffusive
\begin{align}\label{eq:subdiff}
    \omega = -i k^4 \frac{\sigma_{\rm d}}{\dow n/\dow\mu + n^2/B'},
\end{align} 
where $B'=B+2\frac{d-1}{d}G$.
Such subdiffusive modes are characteristic of fracton hydrodynamics. Importantly, here subdiffusion arises without spontaneously broken crystal-dipole symmetry, whereas there are no analogous momentum-conserving phases without spontaneously broken spatial-dipole symmetry~\cite{Glorioso:2023chm, Jensen:2022ibn}. This mode primarily transports charge via $\delta\mu$ fluctuations, but gets mixed with bulk elastic fluctuations $\delta\kappa^i_{~i}=n/B'\delta\mu+\cO(k)$ in a state with nonzero charge density. At higher-orders in $k$, the subdiffusive mode also mixes with $\delta T$, $\delta u^i$, and $\delta\mu_I$ fluctuations; see \cref{app:mode-analysis}.


The constitutive relations for p-wave and s-wave phases additionally depend on $f^{I}_{~J}=\dow p/\dow\xi_{I}^{~J}$. Following through the second law analysis given in \cref{app:second-law}, we find that the crystal-dipole current in the p-wave phase gets a superflow contribution $J^{\mu I}\sim f^{IJ} e_J^\mu$ while the monopole current remains the same. Whereas, in the s-wave phase, the monopole current also gets a superflow contribution $J^\mu \sim -\nabla_\nu\!\lb f^{IJ}e_J^\nu\rb e_I^\mu$. We also find the Josephson equations for the respective Goldstones, $u^\mu \xi_{I\mu} \sim \mu_I$ and $v^\mu\xi_\mu\sim\mu/(u\cdot v)$. The details of the constitutive relations appear in \cref{app:second-law}.

The fluid sound and crystal phonon/diffusion modes remain unaffected in these phases. Whereas, the subdiffusive mode \eqref{eq:subdiff} is affected qualitatively similarly to spatial-dipole superfluids without energy-momentum conservation~\cite{Stahl:2023prt}. In the s-wave phase, the subdiffusive mode gets replaced by a magnon-like propagating mode $\omega\sim\pm k^2$. In the p-wave phase, however, it is replaced by a transverse diffusion mode $\omega\sim-ik^2$ and either two longitudinal diffusion modes $\omega\sim -ik^2$ or magnon-like modes $\omega \sim\pm k^2$ depending on the parameters of the model. The details are included in a supplementary Mathematica notebook~\cite{nb-hydro}.

This spectrum is quite different from the analogous momentum-conserving phases with spatial-dipole symmetry. For instance, the fluid sound mode in the spatial-dipole p-wave phase shows magnon-like dispersion. See a detailed comparison of the mode spectrum in \cref{tab:modes}. Furthermore, the spatial-dipole p-wave phase does not admit any states with nonzero fluid velocity, while such states are unstable in the spatial-dipole s-wave phase~\cite{Gromov:2020yoc, Grosvenor:2021rrt, Jain:2023nbf, Armas:2023ouk, Jain:2024kri}. One the other hand, finite velocity states are automatically stable in our models on account of boost symmetry \footnote{There will still be finite wavevector instabilities that usually appear in relativistic hydrodynamics and may be treated using similar techniques.}.

\section{Fixed crystals}

One may consider physical setups where fractons are embedded on a rigid crystal and thus the elastic dynamics of $\phi^I$ may be ignored. In this case, $\phi^I$ become fixed background fields and the crystal-multipole symmetries effectively reduce to the previously explored spatial-multipole symmetries. Any boost symmetry of the theory also gets explicitly broken by the fixed reference frame $v^\mu$ provided by the crystal. We can combine the fixed $\phi^I$ with the background metric, connection, and crystal-multipole gauge fields to derive the \emph{Aristotelian background} sources suitable for coupling to spatial-multipole-invariant field theories~\cite{Jain:2021ibh, Bidussi:2021nmp}. A detailed comparison is carried out in \cref{app:aristotle}. In fact, if we assume that $\phi^I$ only appears in the theory implicitly via the Aristotelian sources, we immediately arrive at momentum-conserving field theories invariant under spatial-multipole symmetries. 

Generically, however, we expect a fixed background crystal to act as a momentum sink and endow relaxational dynamics to momentum fluctuations~\cite{Armas:2021vku}. To wit, if we allow for arbitrary dependence on $\phi^I$, the operator $K_I$ in \cref{eq:KI} is not set to zero for a non-dynamical crystal and instead shows up as a source in the energy-momentum conservation as $K_I e^{I\nu}$. Expanding around a fixed background $\phi^I = \delta^I_i x^i$, this contribution relaxes velocity/momentum fluctuations with a characteristic rate $\sigma_\phi/(\epsilon+p)$. In other words, the only low-energy variables in such systems are energy and charge with relevant multipole symmetries, and the relevant Goldstone fields for spontaneous symmetry breaking. Interestingly, without a momentum component, one may consistently consider states without spontaneous symmetry breaking~\cite{Glorioso:2023chm, Stahl:2023prt} and the question of nonzero velocity states does not arise in the first place.

\section{Fracton holography}

Fractons have thus far been theoretically realised in a variety of lattice spin models with exotic interactions. However, such models typically feature UV/IR mixing and it is unclear whether there exists an unambiguous renormalisation group prescription that may be used for a bottom-up derivation of their low-energy effective descriptions; see~\cite{Gorantla:2021bda}. It will be interesting to investigate whether the crystal-multipole-invariant theories discussed here fare any better. That said, since crystal-multipole symmetries can be made consistent with relativity, and can be consistently coupled to gravity, they open up an alternate possibility of using holography or the AdS/CFT correspondence for modelling fractons. We note that a holographic model for subdiffusion in the probe limit, i.e. without dynamical gravity in the bulk, has appeared previously in~\cite{Ganesan:2020wvm}.

\begin{figure}
    \centering
    \includegraphics[width=180px]{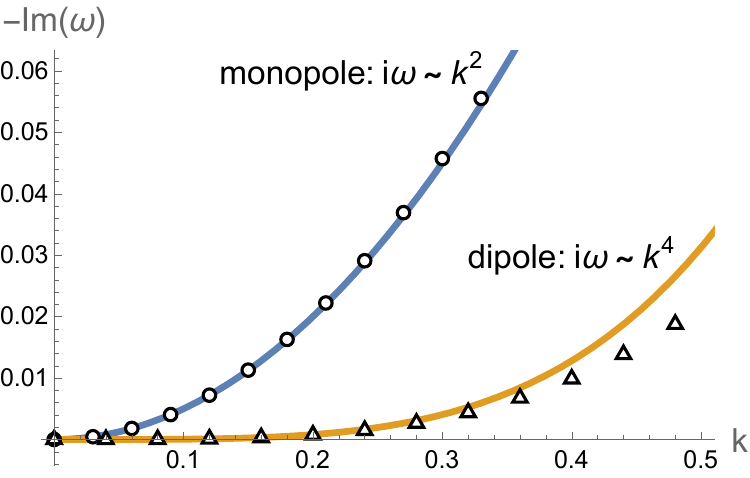}
    \caption{Quasi-normal modes of bulk gauge field fluctuations in the probe limit for monopole and crystal-dipole holographic model. In the absence of $\cA_{I\sM}$, the model exhibits diffusive dispersion $\omega\sim-ik^2$ (circles), which becomes subdiffusive $\omega\sim-ik^4$ when $\cA_{I\sM}$ fluctuations are turned on (triangles).}
    \label{fig:dipole-holography}
\end{figure}

The simplest holographic model combining gravitational and fracton dynamics can be obtained by extending the charged massive gravity model~\cite{Baggioli:2014roa, Alberte:2015isw, Baggioli:2022pyb} to include crystal-dipole gauge fields, i.e.
\begin{align}\label{eq:holofractons}
    S_{\text{bulk}} 
    &= \int_{d+2} \frac{1}{2\kappa}(R - 2\Lambda) - \frac{1}{4g_0} \tilde\cF_{\sM\sN}\tilde\cF^{\sM\sN}
    - \frac{\delta^{IJ}}{4g_1} \cF_{I\sM\sN}\cF_J^{\sM\sN} \nn\\
    &\qquad\qquad 
    - V(X),
\end{align}
Here $R-2\Lambda$ is the Einstein-Hilbert term for the bulk metric $g_{\sM\sN}$, with $\sM,\sN,\ldots$ denoting the $(d+2)$-dimensional bulk indices, $\Lambda=-\half d(d+1)$ is the negative cosmological constant, and $\kappa=8\pi G_{d+2}$ is the gravitational coupling constant. The potential $V$ depends on $X = \half\delta_{IJ}H^{IJ}$, where $H^{IJ} = g^{\sM\sN}\dow_\sM\Phi^I\dow_\sN\Phi^J$ contains the bulk crystal fields $\Phi^I$. Lastly, $\tilde\cF_{\sM\sN}$, $\cF_{I\sM\sN}$ denote the field strengths associated with dynamical monopole and crystal-dipole gauge fields $\cA_\sM$, $\cA_{I\sM}$ in the bulk.

The charged black brane solution in~\cite{Alberte:2015isw} still solves this model with a trivial profile for the dipole gauge field $\cA_{I\sM}=0$. To wit,
\begin{align}
    \df s^2
    &= \frac{1}{z^2}
    \lb -2\df z\,\df t
    - f(z/z_0)\,\df t^2 + \df\vec x^2\rb, \nn\\
    \cA
    &= \mu_0 \frac{z^{d-1}}{z_0^{d-1}} \df t, \nn\\
    \cA_I 
    &= 0, \nn\\
    \Phi^I 
    &= \delta^I_i x^i,
    \label{eq:black-brane-solution}
\end{align}
where $\mu_0$ and $z_0$ are constant parameters. In these coordinates, the horizon resides at $z=z_0$, while the boundary at $z=0$. The blackening factor is given by
\begin{align}
    f(z)
    &= 1-\frac{z^{d+1}}{z_0^{d+1}}
    + \kappa\,\frac{z^{d+1}}{z_0^{d+1}} 
    \left(\frac{z^{d-1}}{z_0^{d-1}}-1\right) 
    \frac{d-1}{d} z_0^2\mu_0^2
    \nn\\
    &\qquad
    - \kappa\,z^{d+1}\frac{2}{d}\int_z^{z_0} \df z 
    \frac{V(X_0)}{z^{d+2}},
\end{align}
where $X_0 = d/2\,z^2$. The Hawking temperature of the black brane is given by
\begin{align}
    T = -\frac{\hbar}{4\pi k_B}f'(z_0).
\end{align}
In the dual boundary theory, this solution corresponds to an equilibrium state with zero crystal-dipole chemical potential. However, the dynamics of this model differ significantly from that of~\cite{Alberte:2015isw} due to fluctuations of $\cA_{I\sM}$. 

In particular, we find that in the probe limit, linearised fluctuations of $\cA_{t}$ and $\cA_{xt}$ decouple from the remaining degrees of freedom; see \cref{app:holography} for more details. The resulting equations can be solved with Dirichlet boundary conditions, resulting in a subdiffusive quasinormal mode
\begin{align}
    \omega = -ik^4\, \frac{g_0}{g_1}D_n + \cO(k^6),\quad 
    D_n = \frac{d+1}{d-1}\frac{\hbar}{4\pi k_BT},
    \label{eq:hol-subdiff}
\end{align}
which replaces the diffusive mode $\omega = -ik^2D_n$ found in ordinary charged holography; see \cref{fig:dipole-holography}.

Interestingly, the holographic model \eqref{eq:holofractons} may describe fractons on a dynamical or fixed background crystal depending on the boundary conditions imposed on the bulk crystal fields $\Phi^I$. However, for the fixed crystal boundary conditions, the momentum fluctuations in the boundary theory are relaxed~\cite{Andrade:2013gsa}, so we expect to find momentum-non-conserving fracton hydrodynamic theories analogous to~\cite{Gromov:2020yoc} with spontaneously unbroken monopole and spatial-dipole symmetries, but with an added conserved energy component. The holographic models for p-wave and s-wave phases, for both dynamical or fixed background crystals, may be obtained by including appropriately charged matter in the bulk along the lines of~\cite{Bhattacharya:2011eea, Baggioli:2022aft}.

\section{Outlook}

In this letter, we have developed fractonic field theories where charged excitations have restricted mobility within a physical crystal, but are permitted to move in absolute space along with the crystal. This is realised through a new class of crystal-multipole symmetries that require the conservation of relevant multipole moments relative to the crystal. We discussed how to gauge these symmetries, giving rise to gapless crystal-multipole gauge theories that are the analogues of tensor gauge theories that arise in the context of spatial-multipole symmetries. We also discussed spontaneously broken crystal-multipole symmetries, giving rise to exotic supersolid phases with mobility restrictions.

The most distinctive aspect of crystal-multipole symmetries is that, unlike their spatial-multipole counterparts, they feature gauge-invariant spatial momenta and thus are compatible with boost symmetries. We exploited this to construct, for the first time, relativistic field theories with fractonic excitations that can be consistently coupled to generic curved gravitational backgrounds. As a result of boost-invariance, these theories also allow for stable hydrodynamic states with nonzero fluid velocity that are forbidden by spatial-multipole symmetries. This has also allowed us to propose a new class of holographic models for studying fractons in the context of AdS/CFT duality. While we have focused on relativistic theories in this work for concreteness, the construction follows analogously for non-relativistic crystal-multipole field theories invariant under Galilean boosts and are left for future work.

We have primarily focused on field theories with crystal-dipole symmetry, where charged excitations are only allowed to move within the crystal in neutral bound states with fixed dipole moment. This construction can easily be extended to higher crystal-multipole symmetries leading to stronger mobility restrictions on charged excitations along the lines of spatial-multipole symmetries in~\cite{Gromov:2018nbv}. We have briefly discussed these extensions and their gauging in the appendix. The simplest one introduces a conserved trace crystal-quadrupole moment $\rmQ_{(2)}$, such that $[\rmP_I,\rmQ_{(2)}] = i\delta_{IJ} \rmQ^{J}$, which restricts bound states to only move perpendicular to the direction of their crystal-dipole moment. 
It will be interesting to explore field theories featuring crystal-multipole symmetries and their low-energy phases in more detail in the future.

We have constructed simple hydrodynamic models with crystal-dipole symmetry and showed how their low-energy behaviour is qualitatively distinct from momentum-conserving hydrodynamics with spatial-dipole symmetry studied previously. However, we have only included a handful of transport parameters to draw out the qualitative features of the mode spectrum. It will be interesting to revisit these models in more detail with the complete characterisation of dissipative and non-dissipative transport along the lines of~\cite{Jain:2023nbf, Armas:2023ouk, Jain:2024kri}. It will also be interesting to explore other extensions of crystal-multiple symmetries such as discrete or non-Abelian internal symmetries, discrete crystalline lattices, or combine these with recent generalised notions of categorical symmetries~\cite{Baez:2010ya, Gaiotto:2014kfa}. 

Furthermore, since the subdiffusive mode in \eqref{eq:subdiff} is carried collectively by charge and elastic excitations, 
the subdiffusion coefficient is sensitive to the elastic moduli of the material on which fractons are lodged. This can be used as an explicit experimental test of the theoretical framework presented in this work. It will also be interesting to explore the fully-backreacted quasinormal mode spectrum of the holographic model in \eqref{eq:holofractons} and test how its predictions for the subdiffusion constant differ from the model with spatial-dipole symmetry in~\cite{Ganesan:2020wvm}.

Real world physical systems feature boost invariance, either Galilean or relativistic. Therefore, fractonic excitations in such systems cannot have mobility restrictions in absolute space and are likely to found lodged to tangible physical materials. By accounting for positional dynamics of the underlying material, and in the process reconciling fractons with spacetime symmetries, we hope that the formalism developed in this letter will prove more robust for understanding this exotic phase of matter and potentially aid in its experimental realisation.

\vspace{1em}

\begin{acknowledgments}
The author would like to thank Jay Armas, Kristan Jensen, Pavel Kovtun, and Eric Mefford for helpful discussions. This work is partly supported by the Dutch Institute for Emergent Phenomena (DIEP) cluster at the University of Amsterdam. Part of this project was carried out during the ``Hydrodynamics at All Scales'' workshop at the Nordic Institute for Theoretical Physics (NORDITA), Stockholm.
\end{acknowledgments}

\appendix
\renewcommand\thefigure{A\arabic{figure}}    
\setcounter{figure}{0} 
\renewcommand{\theequation}{A\arabic{equation}}
\setcounter{equation}{0}

\section{Gauging crystal-dipole symmetry algebra}
\label{app:crystal-dipole}

In this appendix, we discuss the gauging procedure for relativistic crystal-dipole symmetry algebra. The analogous discussion for Galilean systems, or ones without boost symmetry altogether, is straightforward. We introduce the symmetry generators
\begin{align}
    \rmP_{\alpha}:&~\text{spacetime translations}, \nn\\
    \rmM_{\alpha\beta}:&~\text{spacetime Lorentz transformations}, \nn\\
    \rmP_I:&~\text{crystal-translations}, \nn\\
    \rmM_{IJ}:&~\text{crystal-rotations}, \nn\\
    \rmQ:&~\text{U(1) monopole transformations}, \nn\\
    \rmQ^I:&~\text{U(1) crystal-dipole transformations},
\end{align}
where $\rmM_{(\alpha\beta)}=\rmM_{(IJ)}=0$. We may drop $\rmM_{IJ}$ for describing anisotropic crystals. The non-trivial commutation relations in flat spacetime are given as
\begin{align}\label{eq:full-algebra}
    [\rmM_{\alpha\beta},\rmM_{\gamma\delta}] 
    &= 2i\lb\eta_{\gamma[\alpha}\rmM_{\beta]\delta}
    - \eta_{\delta[\alpha}\rmM_{\beta]\gamma} \rb, \nn\\
    [\rmM_{\alpha\beta},\rmP_{\gamma}] 
    &= 2i\eta_{\gamma[\alpha}\rmP_{\beta]}, \nn\\
    [\rmM_{IJ},\rmM_{KL}] 
    &= 2i\lb\delta_{K[I}\rmM_{J]L}
    - \delta_{L[I}\rmM_{J]K} \rb, \nn\\
    [\rmM_{IJ},\rmP_{K}] 
    &= 2i\delta_{K[I}\rmP_{J]}, \nn\\
    [\rmM_{IJ},\rmQ_K] 
    &= 2i\delta_{K[I}\rmQ_{J]}, \nn\\
    [\rmP_I,\rmQ^J] 
    &= i\delta^J_I\rmQ.
\end{align}
Infinitesimal global symmetry transformations may be parametrised by a set of symmetry parameters
\begin{align}
    \scX = \Big(~\chi^\mu~,~
    \Omega^{\alpha}_{~\beta}~,~
    a^I~,~\Omega^{I}_{~J}~,~
    \Lambda~,~\Lambda_I~\Big),
\end{align}
where $\delta^{\beta(\gamma}\Omega^{\alpha)}_{~\beta}=\delta^{J(K}\Omega^{I)}_{~J}=0$. The symmetry variation $\delta_{\scX}$ is expressed in the basis of generators as
\begin{align}\label{eq:variation-flat}
    \delta_{\scX}
    &= -i\chi^\mu \delta_\mu^\alpha \rmP_\alpha
    - \frac{i}{2}\Omega^{\alpha}_{~\beta}    \delta^{\beta\gamma}\rmM_{\gamma\alpha} \nn\\
    &~~
    - ia^I\rmP_I
    - \frac{i}{2}\Omega^{I}_{~J} \delta^{JK}\rmM_{KI}
    - i\Lambda\rmQ 
    - i\Lambda_I\rmQ^I.
\end{align}
Given the commutation relations in \cref{eq:full-algebra} and requiring that the infinitesimal symmetry transformations form an algebra $[\delta_{\scX'},\delta_{\scX}] = \delta_{[\scX',\scX]}$, we are led to the commutator 
\begin{equation}
    [\scX',\scX] = \delta_{\scX'}\scX = - \delta_{\scX}\scX',
\end{equation}
with the transformation rules of the symmetry parameters themselves given by
\begin{align}
    \delta_{\scX'}\chi^\mu 
    &= \lie_{\chi'}\chi^\mu, \nn\\
    \delta_{\scX'}\Omega^{\alpha}_{~\beta}
    &= \lie_{\chi'}\Omega^{\alpha}_{~\beta}
    - \lie_{\chi}\Omega'^{\alpha}_{~\beta}
    + \Omega^{\alpha}_{~\gamma}\Omega'^{\gamma}_{~\beta}
    - \Omega'^{\alpha}_{~~\gamma}\Omega^{\gamma}_{~\beta}
    , \nn\\
    \delta_{\scX'}a^I 
    &= \lie_{\chi'}a^I - \lie_\chi a'^I
    - \Omega'^{I}_{~J} a^J
    + \Omega^{I}_{~J} a'^J
    , \nn\\
    \delta_{\scX'}\Omega^{I}_{~J}
    &= \lie_{\chi'}\Omega^{I}_{~J}
    - \lie_{\chi}\Omega'^{I}_{~J}
    + \Omega^{I}_{~K}\Omega'^{K}_{~J}
    - \Omega'^{I}_{~K}\Omega^{K}_{~J}
    , \nn\\
    \delta_{\scX'}\Lambda 
    &= \lie_{\chi'}\Lambda - \lie_\chi\Lambda'
    - a'^I \Lambda_I + a^I\Lambda'_I, \nn\\
    \delta_{\scX'}\Lambda_I
    &= \lie_{\chi'}\Lambda_I - \lie_\chi\Lambda'_I
    + \Lambda_J\Omega'^J_{~I}
    - \Lambda'_J\Omega^J_{~I}.
\end{align}
These are the standard transformation rules for symmetry parameters, besides the additional terms in the transformation of $\Lambda$ implying that we can generate a monopole transformation by commuting crystal-translations and crystal-dipole transformations.

To probe the conserved currents associated with the symmetry generators, it is convenient to introduce a set of background fields
\begin{align}
    \tau^\alpha_\mu:&~\text{spacetime vielbein}, \nn\\
    \omega^{\alpha}_{~\beta\mu}:&~\text{spacetime spin-connection}, \nn\\
    \tau^I_\mu:&~\text{crystal vielbein}, \nn\\
    \omega^I_{~J\mu}:&~\text{crystal spin-connection}, \nn\\
    A_\mu:&~\text{U(1) monopole gauge field}, \nn\\
    A_{I\mu}:&~\text{U(1) crystal-dipole gauge field},
\end{align}
such that $\delta^{\beta(\gamma}\omega^{\alpha)}_{~\beta\mu}=\delta^{J(K}\omega^{I)}_{~J\mu}=0$. The spacetime vielbein $\tau^\alpha_\mu$ is required to invertible, with the inverse denoted as $\tau^\mu_\alpha$, but we do not require invertibility for $\tau^I_\mu$. In the presence of background fields, the definition of the symmetry variation $\delta_{\scX}$ modifies to
\begin{align}
    \delta_{\scX}
    &= -i\chi^\mu \tau_\mu^\alpha \rmP_\alpha
    - \frac{i}{2}\lb\Omega^{\alpha}_{~\beta} + \chi^\mu \omega^{\alpha}_{~\beta\mu}\rb
    \delta^{\beta\gamma}\rmM_{\gamma\alpha} \nn\\
    &~~
    - i\lb a^I + \chi^\mu\tau_\mu^I\rb \rmP_I
    - \frac{i}{2}\lb \Omega^{I}_{~J} + \chi^\mu \omega^{I}_{~J\mu}\rb 
    \delta^{JK}\rmM_{KI} \nn\\
    &~~
    - i\lb \Lambda + \chi^\mu A_\mu \rb\rmQ 
    - i\lb \Lambda_I + \chi^\mu A_{I\mu} \rb\rmQ^I.
\end{align}
We note that spacetime translations do not commute in the presence of background sources, i.e.
\begin{align}
    &[\rmP_\alpha,\rmP_\beta] \tau^\alpha_\mu\tau^\beta_\nu
    = -iT^\alpha_{~\mu\nu} \rmP_\alpha
    - \frac{i}{2} R^{\alpha}_{~\beta\mu\nu} \delta^{\beta\gamma}\rmM_{\gamma\alpha} \nn\\
    &\hspace{-0.6em}
    - iT^I_{~\mu\nu} \rmP_I
    - \frac{i}{2} R^I_{~J\mu\nu}\delta^{JK} \rmM_{KI}
    - iF_{\mu\nu}\rmQ 
    - iF_{I\mu\nu} \rmQ^I,
\end{align}
where we have defined various torsions, curvatures, and field strength tensors
\begin{align}\label{eq:invariants}
    T^\alpha_{~\mu\nu}
    &= 2\dow_{[\mu}\tau^\alpha_{\nu]} + 2\omega^{\alpha}_{~\beta[\mu}\tau_{\nu]}^\beta, \nn\\
    R^{\alpha}_{~\beta\mu\nu}
    &= 2\dow_{[\mu}\omega^{\alpha}_{~\beta\nu]} 
    + 2\omega^{\alpha}_{~\gamma[\mu}\omega^\gamma_{~\beta\nu]}, \nn\\
    T^I_{~\mu\nu}
    &= 2\dow_{[\mu}\tau^I_{\nu]} + 2\omega^{I}_{~J[\mu}\tau_{\nu]}^J, \nn\\
    R^{I}_{~J\mu\nu}
    &= 2\dow_{[\mu}\omega^{I}_{~J\nu]} + 2\omega^I_{~K[\mu}\omega^K_{~J\nu]}, \nn\\
    F_{\mu\nu}
    &= 2\dow_{[\mu}A_{\nu]} + 2\tau^I_{[\mu} A_{I\nu]}, \nn\\
    F_{I\mu\nu}
    &= 2\dow_{[\mu}A_{I\nu]} - 2\omega^J_{~I[\mu}A_{J\nu]},
\end{align}
Note the unconventional definition of $F_{\mu\nu}$ due to the non-trivial dipole structure of the theory. Imposing $[\delta_{\scX'},\delta_{\scX}] = \delta_{[\scX',\scX]}$, we may derive the symmetry transformations of the background fields
\begin{align}\label{eq:background-symmetries}
    \delta_{\scX} \tau^\alpha_\mu 
    &= \lie_\chi \tau^\alpha_\mu 
    - \Omega^{\alpha}_{~\beta}\tau^\beta_\mu \nn\\
    &= \chi^\nu T^\alpha_{~\mu\nu}
    + \nabla_\mu (\chi^\nu e_\nu^\alpha)
    - \lb\Omega^{\alpha}_{~\beta} + \chi^\nu \omega^\alpha_{~\beta\nu}\rb\tau^\beta_\mu, \nn\\[0.5em]
    \delta_{\scX} \omega^{\alpha}_{~\beta\mu}
    &= \lie_\chi \omega^{\alpha}_{~\beta\mu}
    + \dow_\mu \Omega^{\alpha}_{~\beta}
    - \Omega^{\alpha}_{~\gamma}\omega^{\gamma}_{~\beta\mu}
    + \omega^{\alpha}_{~\gamma\mu}\Omega^{\gamma}_{~\beta} \nn\\
    &= \chi^\nu R^{\alpha}_{~\beta\nu\mu}
    + \nabla_\mu\!\lb\Omega^{\alpha}_{~\beta} + \chi^\nu \omega^\alpha_{~\beta\nu}\rb, \nn\\[0.5em]
    \delta_{\scX}\tau^I_\mu 
    &= \lie_\chi\tau^I_\mu 
    + \dow_\mu a^I
    + \omega^I_{~J\mu} a^J
    - \Omega^I_{~J} \tau^J_\mu \nn\\
    &= \chi^\nu T^I_{~\nu\mu}
    + \nabla_\mu\!\lb a^I + \chi^\nu \tau_\nu^I\rb \nn\\
    &\qquad 
    - \lb\Omega^{I}_{~J} + \chi^\nu \omega^I_{~J\nu}\rb \tau^J_\mu, \nn\\[0.5em]
    \delta_{\scX} \omega^I_{~J\mu}
    &= \lie_\chi \omega^I_{~J\mu}
    + \dow_\mu \Omega^I_{~J}
    - \Omega^{I}_{~K}\omega^K_{~J\mu}
    + \omega^I_{~K\mu}\Omega^{K}_{~J} \nn\\
    &= \chi^\nu R^{I}_{~J\nu\mu}
    + \nabla_\mu\!\lb\Omega^{I}_{~J} + \chi^\nu \omega^I_{~J\nu}\rb, \nn\\[0.5em]
    \delta_\scX A_\mu
    &= \lie_\chi A_\mu
    + \dow_\mu\Lambda 
    - a^I A_{I\mu}
    + \Lambda_I \tau^I_\mu \nn\\
    &= \chi^\nu F_{\nu\mu}
    + \dow_\mu\!\lb\Lambda + \chi^\nu A_\nu\rb \nn\\
    &\qquad 
    - \lb a^I + \chi^\nu \tau^I_\nu \rb A_{I\mu}
    + \lb \Lambda_I + \chi^\nu A_{I\nu}\rb \tau^I_\mu, \nn\\[0.5em]
    \delta_{\scX} A_{I\mu}
    &= \lie_\chi A_{I\mu} 
    + \dow_\mu\Lambda_I
    - \Lambda_J \omega^J_{~I\mu} 
    + A_{J\mu}\Omega^J_{~I} \nn\\
    &= \chi^\nu F_{I\nu\mu} 
    + \nabla_\mu\!\lb\Lambda_I + \chi^\nu A_{I\nu}\rb \nn\\
    &\qquad 
    + A_{J\mu}\lb\Omega^J_{~I} + \chi^\nu \omega^J_{~I\nu}\rb,
\end{align}
where $\nabla_\mu$ is the covariant derivative associated with spin connections $\omega^\alpha_{~\beta\mu}$, $\omega^I_{~J\mu}$, and the spacetime connection
\begin{align}
    \Gamma_{\mu\nu}^\lambda
    &= \tau^\lambda_\alpha\dow_\mu \tau^\alpha_\nu 
    + \tau^\lambda_\alpha\omega^\alpha_{~\beta\mu}\tau^\beta_\nu,
\end{align}
defined such that $\nabla_\mu\tau^\alpha_\nu = 0$.
\Cref{eq:background-symmetries} gives rise to the transformation rules of gauge fields in \cref{eq:gauge-trans} when $\tau^I_\mu$ and $\omega^I_{~J\mu}$ are swiched off.
The tensors in \cref{eq:invariants} are invariant under monopole transformations and transform covariantly under diffeomorphisms, spacetime Lorentz transformations, and crystal-rotations. On the other hand, crystal-translations and crystal-dipole transformations act non-trivially on
\begin{align}
    T^I_{~\mu\nu}
    &\to T^I_{~\mu\nu} + R^I_{~J\mu\nu} a^J, \nn\\
    F_{\mu\nu}
    &\to  F_{\mu\nu} -  a^I F_{I\mu\nu}
    + \Lambda_I T^I_{~\mu\nu} , \nn\\
    F_{I\mu\nu}
    &\to F_{I\mu\nu}
    - \Lambda_J R^J_{~I\mu\nu}.
\end{align}

So far, we have treated crystal translations and rotations as abstract symmetries. However, they are fundamentally different from the other symmetries in that they are tied exclusively to the dynamical crystal fields that transform under infinitesimal transformations as
\begin{align}
    \delta_\scX \phi^I = \lie_\chi \phi^I + a^I - \Omega^{IJ}\phi_J.
\end{align}
Thus the crystal vielbein $\tau^I_\mu$ and crystal spin-connection $\omega^I_{~J\mu}$ basically just keep track of the derivatives of $\phi^I$ via the covariant crystal frame fields
\begin{align}
    e^I_\mu = \dow_\mu \phi^I - \tau^I_\mu + \omega^I_{~J\mu} \phi^J.
\end{align}
We can also use $\phi^I$ to define versions of $T^I_{~\mu\nu}$ and $F_{\mu\nu}$ that are invariant under crystal-translations
\begin{align}
    \tilde T^I_{~\mu\nu} &= T^{I}_{~\mu\nu} - R^I_{~J\mu\nu}\phi^J, \nn\\
    \tilde F_{\mu\nu} &= F_{\mu\nu} + \phi^I F_{I\mu\nu}.
\end{align}
Note that the field strengths $\tilde F_{\mu\nu}$ and $F_{I\mu\nu}$ are also invariant under crystal-dipole transformations when the background crystal-torsion $T^{I}_{~\mu\nu}$ and crystal-curvature $R^I_{~J\mu\nu}$ are switched off.

In the following, we will exclusively work with torsionless ambient spacetimes, where the spin-connection $\omega^\alpha_{~\beta\mu}$ is fixed in terms of the vielbein $\tau^\alpha_\mu$ as
\begin{align}
    \omega^{\alpha}_{~\beta\mu}
    &= \tau^{\alpha\nu} \dow_{[\mu} \tau_{\beta\nu]}
    - \tau^{\nu}_\beta \dow_{[\mu} \tau^\alpha_{\nu]}
    - \tau^{\alpha\rho} \tau^{\nu}_\beta \tau_{\gamma\mu} \dow_{[\rho}\tau_{\nu]}^\gamma,
\end{align}
leading to $T^\alpha_{~\mu\nu}=0$.
In this case, all the Lorentz-invariant information in $\tau^\alpha_\mu$ is captured by the spacetime metric $g_{\mu\nu} = \tau^\alpha_\mu \tau_{\alpha\nu}$, which transforms as simply
\begin{align}
    \delta_\scX g_{\mu\nu}
    &= \lie_\chi g_{\mu\nu}
    = 2\nabla_{(\mu}\chi_{\nu)}.
\end{align}
Note that a similar construction does not apply to the crystal spin-connection $\omega^I_{~J\mu}$ because $\tau^I_\mu$ is not gauge-invariant and may not be invertible either.

\section{Ward identities}

Given an effective action $S$ of a field theory coupled to torsionless spacetime background, the variations with respect to background sources and the crystal fields $\phi^I$ can be parametrised as
\begin{align}\label{eq:action-variation-app}
    \delta S
    &=\int_{d+1} \half T^{\mu\nu} \delta g_{\mu\nu} \nn\\
    &~~
    + r^{\mu}_I \lb \delta \tau^I_\mu - \phi^J\delta \omega^I_{~J\mu}\rb
    + \half \Sigma^{\mu I}_{~~J}\delta \omega^J_{~I\mu}
    \nn\\
    &~~
    + J^\mu \lb \delta A_\mu + \phi^I \delta A_{I\mu}\rb 
    + J^{I\mu} \delta A_{I\mu}
    + K_I \delta \phi^I.
\end{align}
Requiring the effective action to be invariant under the spacetime diffeomorphisms, U(1) monopole, and U(1) crystal-dipole transformations, given in \cref{eq:background-symmetries}, leads to the Ward identities 
\begin{align}
    \nabla_\mu T^{\mu\nu}
    &= \tilde F^{\nu\rho} J_\rho 
    + F_I^{\nu\rho} J^I_\rho \nn\\
    &\qquad 
    + \tilde T^{I\nu\rho} r_{I\rho}
    + \half R^{I}_{~J}{}^{\nu\rho} \Sigma_{\rho}{}^J_{~I}
    + K_Ie^{I\nu}, \nn\\
    \nabla_\mu J^\mu 
    &= 0, \nn\\
    \nabla_\mu J^{I\mu}
    &= - e^I_\mu J^\mu,
\end{align}
They give rise to the conservation equations in \cref{eq:cons-free} when $T^{I}_{~\mu\nu}$ and $R^I_{~J\mu\nu}$ are switched off and $\phi^I$ is taken onshell.
The crystal-translation and crystal-rotation symmetries, on the other hand, yield
\begin{align}
    \nabla_\mu r^\mu_I 
    &= - A_{I\mu} J^\mu
    + K_I, \nn\\
    \nabla_\mu \Sigma^{\mu IJ}
    &= - 2r^{[I\mu} e^{J]}_\mu 
    + 2J^{[I\mu} A^{J]}_\mu,
\end{align}
which essentially fix the form of the equation of motion of the crystal fields $\phi^I$, i.e. $K_I = \nabla_\mu r^\mu_I +  A_{I\mu} J^\mu =0$.
For an isotropic crystal, we can further eliminate the antisymmetric spatial part of $r^\mu_I$ using the crystal-rotation Ward identity.

\section{Crystal-multipole symmetries}
\label{app:crystal-multipole}

The symmetry algebra can be easily extended to arbitrary higher crystal-$2^n$-pole symmetries with generators
\begin{equation}
    \rmQ^{I_1\ldots I_r}_{(r)}:~\text{U(1) crystal-$2^r$-pole transformations},
\end{equation}
for $0\leq r\leq n$, and the non-trivial commutators
\begin{equation}
    [\rmP_I,\rmQ^{J_1\ldots J_r}_{(r)}]
    = ir\delta^{(J_1}_I \rmQ_{(r-1)}^{J_2\ldots J_r)}.
\end{equation}
For example, the simplest such extension consists of the trace crystal-quadrupole symmetry, $\rmQ_{(2)}^{IJ}\sim 2/d\,\delta^{IJ}\rmQ_{(2)}$, satisfying $[\rmP_I,\rmQ_{(2)}] = i\delta_{IJ} \rmQ_{(1)}^{J}$. The infinitesimal symmetry variation $\delta_\scX$ can be extended to include crystal-multipole variations as
\begin{equation}
    \delta_\scX \sim 
    -i\sum_{r=0}^n \frac{1}{r!}\Lambda^{(r)}_{I_1\ldots I_r}
    \rmQ^{I_1\ldots I_r}_{(r)},
\end{equation}
with the symmetry parameters $\Lambda^{(r)}_{I_1\ldots I_r}$. Imposing the commutation relations $[\delta_{\scX'},\delta_{\scX}] = \delta_{[\scX',\scX]}$, we find the transformation rules of the symmetry parameters
\begin{align}
    \delta_{\scX'} \Lambda^{(r)}_{I_1\ldots I_r}
    &= \lie_{\chi'}\Lambda^{(r)}_{I_1\ldots I_r} 
    - \lie_\chi\Lambda'^{(r)}_{I_1\ldots I_r}
    \nn\\
    &\qquad 
    - a'^J \Lambda^{(r+1)}_{JI_1\ldots I_r}
    + a^J \Lambda'^{(r+1)}_{JI_1\ldots I_r}, 
\end{align}
where crystal-translations mix them with the next higher multipole-order transformations.

We can gauge the crystal-multipole symmetries by introducing the background fields
\begin{align}
    A^{(r)}_{I_1\ldots I_r\mu}:&~\text{U(1) crystal-$2^r$-pole gauge field},
\end{align}
used to generalise the symmetry variation to
\begin{equation}
    \delta_\scX \sim 
    -i\sum_{r=0}^n \frac{1}{r!}
    \lb \Lambda^{(r)}_{I_1\ldots I_r} + \chi^\mu A^{(r)}_{I_1\ldots I_r\mu} \rb
    \rmQ^{I_1\ldots I_r}_{(r)}.
\end{equation}
In the presence of these, the commutator of spacetime translations gets additional contributions
\begin{align}
    [\rmP_\alpha,\rmP_\beta] \tau^\alpha_\mu\tau^\beta_\nu
    \sim - \frac{i}{r!}F_{I_1\ldots I_r\mu\nu}^{(r)} \rmQ^{I_1\ldots I_r}_{(r)},
\end{align}
where the field strengths are defined as
\begin{align}
    F^{(r)}_{I_1\ldots I_r\mu\nu} 
    &= 2\dow_{[\mu}A^{(r)}_{I_1\ldots I_r\nu]}
    - 2r\omega^{J}_{~[I_r\mu} A^{(r)}_{I_1\ldots I_{r-1}]J\mu} \nn\\
    &\qquad 
    + 2\tau^{J}_{[\mu} A^{(r+1)}_{JI_1\ldots I_{r}\nu]}.
\end{align}
Requiring $[\delta_{\scX'},\delta_{\scX}] = \delta_{[\scX',\scX]}$, we may derive the transformation rules of the crystal-multipole gauge fields
\begin{align}
    \delta_\scX & A^{(r)}_{I_1\ldots I_r\mu}\nn\\
    &= \lie_\chi A^{(r)}_{I_1\ldots I_r\mu} 
    + \dow_\mu \Lambda^{(r)}_{I_1\ldots I_r}
    + r\omega^J_{~[I_{r}\mu} \Lambda^{(r)}_{I_1\ldots I_{r-1}]}
    \nn\\
    &\qquad 
    + A^{(r)}_{J[I_2\ldots I_r\mu} \Omega^J_{~I_{1}]}
    \nn\\
    &\qquad 
    - a^{J} A^{(r+1)}_{JI_1\ldots I_r\mu}
    + \Lambda^{(r+1)}_{JI_1\ldots I_r} \tau^J_\mu \nn\\
    &= \chi^\nu F^{(r)}_{I_1\ldots I_r\nu\mu} 
    + \nabla_\mu\!\lb \Lambda^{(r)}_{I_1\ldots I_r} + \chi^\nu A^{(r)}_{I_1\ldots I_r\nu} \rb
    \nn\\
    &\qquad 
    + A^{(r)}_{J[I_2\ldots I_r\mu} \lb \Omega^J_{~I_1]} + \chi^\nu \omega^J_{~I_1]\nu}\rb
    \nn\\
    &\qquad 
    - \lb a^{J} + \chi^\nu \tau_\nu^J \rb A^{(r+1)}_{JI_1\ldots I_r\mu} \nn\\
    &\qquad
    + \lb \Lambda^{(r+1)}_{JI_1\ldots I_r} 
    + \chi^\nu A^{(r+1)}_{JI_1\ldots I_r\nu} \rb \tau^J_\mu.
\end{align}

Consider the effective action $S$ of a field theory invariant under crystal-multipole transformations. The variation in \cref{eq:action-variation-app} is extended to include variations with respect to the background crystal-multipole gauge fields
\begin{align}
    \delta S
    &\sim \int_{d+1}
    \sum_{r=0}^n \frac{1}{r!} J^{\mu I_1\ldots I_r}_{(r)}
    \Bigg( \delta A_{I_1\ldots I_r\mu}^{(r)} \nn\\
    &\qquad\qquad
    + \sum_{s=1}^{n-r}\frac{1}{s!}\phi^{J_1}\ldots \phi^{J_s} \delta A^{(r+s)}_{J_1\ldots J_{s}I_1\ldots I_r\mu}
    \Bigg).
\end{align}
The choice of the parametrisation here is such that the currents $J^{\mu I_1\ldots I_r}_{(r)}$ are invariant under crystal-translations. Demanding the action to be invariant under spacetime diffeomorphisms and crystal-multipole symmetries lead to the energy-momentum and multipole conservation equations for these currents
\begin{align}
    \nabla_\mu T^{\mu\nu}
    &= \sum_{r=0}^n\frac{1}{r!} \tilde F^{(r)}_{I_1\ldots I_r}{}^{\nu\rho} 
    J_{(r)\rho}{}^{I_1\ldots I_r} \nn\\
    &\qquad
    + \tilde T^{I\nu\rho} r_{I\rho}
    + \half R^{I}_{~J}{}^{\nu\rho} \Sigma_{\rho}{}^J_{~I} 
    + K_Ie^{I\nu}, \nn\\
    \nabla_\mu J^{\mu I_1\ldots I_r}_{(r)}
    &= -r e^{(I_1}_\mu
    J^{\mu I_2\ldots I_{r})}_{(r-1)},
\end{align}
where we have defined the crystal-translation-invariant combinations of field strengths
\begin{align}
    \tilde F^{(r)}_{I_1\ldots I_r\mu\nu}
    &= F^{(r)}_{I_1\ldots I_r\mu\nu} \nn\\
    &~~
    + \sum_{s=1}^{n-r}\frac{1}{s!}
    \phi^{J_1}\ldots \phi^{J_s} 
    F^{(r+s)}_{J_1\ldots J_sI_1\ldots I_r\mu\nu}.
\end{align}

Similar to the discussion in the main text for crystal-dipole symmetries, we can construct effective field theories invariant under crystal-multipole symmetries. As an example, consider a complex scalar field theory invariant under U(1) monopole $\Lambda^{(0)}$, crystal-dipole $\Lambda^{(1)}_I$, and trace crystal-quadrupole $\Lambda^{(2)}_{IJ} = \Lambda^{(2)}\delta_{IJ}$ symmetry, with the Lagrangian
\begin{equation}
    {\cal L}_\Psi 
    = |\Df_v \Psi|^2
    - \lambda |\rmD_{\{IJ\}}(\Psi,\Psi)|^2
    - V(|\Psi|^2).
\end{equation}
Here $X_{\{IJ\}}= X_{IJ} - \frac1d\,\delta_{IJ}\delta^{KL}X_{KL}$ denotes the traceless combination with respect to $\delta^{IJ}$ (as opposed to $h^{IJ}$). To gauge the trace crystal-quadrupole symmetry, we only need to introduce the trace part of the respective gauge field $A^{(2)}_{IJ\mu} = \delta_{IJ} A^{(2)}_\mu$. Correspondingly, the definition of gauge-covariant derivative gets updated to
\begin{equation}
    \Df_\mu\Psi 
    = \dow_\mu\Psi 
    + i\lb A_\mu^{(0)} + \phi^IA_{I\mu}^{(1)}
    + \half\phi^I\phi^J\delta_{IJ} A_{\mu}^{(2)}
    \rb,
\end{equation}
together with the updated bilocal derivative operator
\begin{align}
    \Df_{IJ}(\Psi,\Psi) 
    &= \Psi\Df_{(I}\Df_{J)} \Psi - \Df_I\Psi\Df_J\Psi \nn\\
    &\hspace{-2em}
    - i\lb e^\mu_{(I} A^{(1)}_{J)\mu}
    + e^\mu_{(I} \delta_{J)K}\phi^K A_{\mu}^{(2)}\rb \Psi^2.
\end{align}
In particular, one may check that the traceless combination $\rmD_{\{IJ\}}(\Psi,\Psi)$ transforms covariantly under local gauge transformations. For theories with trace crystal-quadrupole symmetry, the conservation equations are
\begin{align}
    \nabla_\mu T^{\mu\nu}
    &= \tilde F^{(0)\nu\rho} J_{(0)\rho} 
    + \tilde F^{(1)\nu\rho}_{I} J^I_{(1)\rho} 
    + F^{(2)\nu\rho} J_{(2)\rho} \nn\\
    &\qquad 
    + \tilde T^{I\nu\rho} r_{I\rho}
    + \half R^{I}_{~J}{}^{\nu\rho} \Sigma_{\rho}{}^J_{~I}
    + K_Ie^{I\nu}, \nn\\
    \nabla_\mu J^{\mu}_{(0)}
    &= 0, \nn\\
    \nabla_\mu J^{I\mu}_{(1)}
    &= -  e^{I}_\mu J^{\mu}_{(0)}, \nn\\
    \nabla_\mu J^{\mu}_{(2)}
    &= - \delta_{IJ} e^{I}_\mu J^{J\mu}_{(1)}.
\end{align}
We have identified the trace crystal-quadrupole current $J^{\mu IJ}_{(2)} = 2/d\,\delta^{IJ} J_{(2)}^\mu$, together with
\begin{align}
    \tilde F^{(0)}_{\mu\nu} 
    &= F^{(0)}_{\mu\nu} + \phi^{I} F^{(1)}_{I\mu\nu}
    + \frac{1}{2} \phi^I\phi^J\delta_{IJ} F^{(2)}_{\mu\nu}, \nn\\
    \tilde F^{(1)}_{I\mu\nu}
    &= F^{(1)}_{I\mu\nu} + \delta_{IJ}\phi^{J}F^{(2)}_{\mu\nu}.
\end{align}
The explicit conserved currents can be obtained by varying the effective action with respect to the appropriate background gauge fields.

\section{Details of complex scalar field theory with crystal-dipole symmetry}
\label{app:complex-scalar}

In this appendix, we give the details of conserved currents and equations of motion for crystal-dipole-invariant complex scalar field theory in \cref{eq:complex-scalar-L}.
Let us begin with the variational identities
\begin{align}
    \delta h^{IJ}
    &= 2 e^{(I\mu} \delta e^{J)}_\mu 
    - e^{I\mu}e^{J\nu} \delta g_{\mu\nu}, \nn\\
    \delta v^\mu 
    &= - e^\mu_I v^\nu \delta e^I_\nu 
    + \half v^\mu v^\rho v^\sigma \delta g_{\rho\sigma}, \nn\\
    \delta e^\mu_I 
    &= - \lb h_{IJ} v^\mu v^\nu 
    + e^\mu_J e^\nu_I 
    \rb \delta e^J_\nu
    + v^\mu v^\rho e_I^\sigma \delta g_{\rho\sigma}.
\end{align}
Using these, we can find
\begin{align}
    \delta \Df_v \Psi
    &= \lb 
    \Df_v\Df_I\Psi
    - iv^\mu A_{I\mu} \Psi
    \rb \delta\phi^I 
    \nn\\
    &\qquad 
    + \Df_v\!\lb\delta\Psi -\Df_I\Psi\delta\phi^I\rb
    + \Df_v \Psi \half v^\mu v^\nu \delta g_{\mu\nu} \nn\\
    &\qquad 
    - iv^\mu \Psi \lb \delta A_\mu + \phi^I\delta A_{I\mu} \rb,
    \nn\\[0.5em]
    \delta \Df_I \Psi
    &= \lb 
    \Df_I\Df_J\Psi
    - h_{IJ} \Df_v\Psi\,\Df_v
    - ie^\mu_J A_{I\mu} \Psi\rb \delta\phi^J \nn\\
    &\qquad 
    + \Df_I\!\lb\delta\Psi -\Df_J\Psi\delta\phi^J\rb
    + \Df_v\Psi\, v^\mu e_I^\nu \delta g_{\mu\nu}
    \nn\\
    &\qquad 
    - ie^\mu_I \Psi
    \lb \delta A_\mu + \phi^I\delta A_{I\mu} \rb.
\end{align}
The bilocal space-space and time-space derivative operators for differently charged complex field arguments are defined as
\begin{align}
    \Df_{IJ}(\Psi_1,\Psi_2) 
    &= \frac{q_2}{2q_1}\Psi_2\Df_{(I}\Df_{J)} \Psi_1 
    + \frac{q_1}{2q_2}\Psi_1\Df_{(I}\Df_{J)} \Psi_2 \nn\\
    &\qquad 
    - \Df_{(I}\Psi_1\Df_{J)}\Psi_2
    - i\frac{q_1+q_2}{2} e^{\mu}_{(I} A_{J)\mu} \Psi^2, \nn\\
    \Df_{vI}(\Psi_1,\Psi_2)
    &= \frac{q_2}{2q_1}\Psi_2\{\Df_v,\Df_{I}\}\Psi_1
    - \Df_{I}\Psi_2 \Df_v\Psi_1 \nn\\
    &\qquad 
    - \frac{i}{2}q_2 A_{I\lambda} v^\lambda \Psi^2,
\end{align}
where $\{\circ,\circ\}$ denotes the anticommutator. Using these definitions, we can compute
\begin{widetext}
\begin{align}
    \delta \Df_{IJ}(\Psi,\Psi)
    &= 2\Df_{IJ}\!\lb\Psi,\delta \Psi - \Df_K \Psi \delta\phi^K\rb
    - ie^\mu_{(I} \delta A_{J)\mu} \Psi^2 
    - i\Psi^2 \dow_{(I}
    \Big( e^\mu_{J)}\lb \delta A_\mu + \phi^K\delta A_{K\mu}\rb 
    \Big)
    + \Df_K \Df_{IJ}(\Psi,\Psi) \delta\phi^K
    \nn\\
    &\qquad 
    + \bigg( 2\Df_{v(I}(\Psi,\Psi)
    + \Psi \Df_v\Psi\, \dow_{(I}
    \bigg) \lb 
    v^\mu e_{J)}^\nu\delta g_{\mu\nu}
    - h_{J)K}\Df_v\delta\phi^K
    + \omega_{J)K} \delta\phi^K
    \rb 
    \nn\\
    &\qquad 
    + i\Psi^2 e_{K}^\mu e^\nu_{(I} \lb \dow_{J)}\tilde F_{\mu\nu} 
    +  F_{J)\mu\nu} 
    + \tilde F_{\mu\nu}\dow_{J)}
    \rb \delta\phi^K,
\end{align}
\end{widetext}
where $\omega_{IJ} = 2e^\mu_{[I}e^\nu_{J]}\dow_\mu v_\nu$ is the crystal vorticity tensor. With these in place, we can use the variational formulae in \cref{eq:action-variation} to read off $J^\mu$ and $J^{\mu I}$ given in \cref{eq:scalar-field-currents}, together with $T^{\mu\nu}$ given in \cref{eq:em-tensor}.
We can also use the variational formulae to read off the equations of motion for $\Psi$ and $\phi^I$, parametrised as
\begin{align}
    \delta S &= \int_{d+1} K_\Psi^\dagger\lb \delta\Psi - \Df_I\Psi\delta\phi^I\rb
    + K_\Psi \lb \delta\Psi^\dagger - \Df_I\Psi^\dagger \delta\phi^I\rb \nn\\
    &\qquad\qquad\qquad
    + K_I \delta\phi^I.
\end{align}
For $\Psi$, we find
\begin{align}
    K_\Psi^\dagger
    &= - \nabla_\mu\!\lb v^\mu \Df_v\Psi^\dagger \rb 
    - V'(|\Psi|^2) \Psi^\dagger \nn\\
    &\qquad
    + 4\lambda\Df_{IJ}\Big(\Df^{\dagger IJ}(\Psi,\Psi),\Psi\Big) \nn\\
    &\qquad
    - 2\lambda X_I \Df_J\Big(\Df^{\dagger IJ}(\Psi,\Psi),\Psi\Big) \nn\\
    &\qquad
    - \lambda\lb \Df_I X_J + X_I X_J \rb \Df^{\dagger IJ}(\Psi,\Psi) \Psi,
\end{align}
where $X_I = \nabla_\mu e^\mu_I$, whereas for $\phi^I$ we find
\begin{widetext}
\begin{align}
    K_I
    &= \Df_v\Psi^* \lb \Df_v\Df_I\Psi
    - iv^\mu A_{I\mu} \Psi\rb 
    - \half \dow_I\bigg(
    \lambda \Df^{\dagger}_{JK}(\Psi,\Psi)\Df^{JK}(\Psi,\Psi)
    + V(|\Psi|^2) \bigg)
    \nn\\
    &~~
    - \lambda 
    \Big( h_{IJ}\Df_v
    + 2e^\mu_Ie^\nu_J \nabla_\nu v_\mu
    + h_{JI}\nabla_\nu v^\nu \Big)
    \bigg( 2\Df^{\dagger KJ}(\Psi,\Psi)\Df_{vK}(\Psi,\Psi)
    - \nabla_{\mu}\!\lb e^\mu_K \Df^{\dagger KJ}(\Psi,\Psi)\Psi \Df_v\Psi\rb \bigg) 
    \nn\\
    &~~
    + i\lambda \lb \tilde F_{\mu\nu} \dow'_{J}
    -  F_{J\mu\nu} 
    \rb \lb e_I^\mu e^\nu_K \Df^{\dagger JK}(\Psi,\Psi) \Psi^2 \rb 
    + 2\lambda \nabla_\mu\lb\Df^\dagger_{K(I}(\Psi,\Psi) 
    \Df_{J)}^{~K}(\Psi,\Psi) e^{J\mu}\rb
    + \text{c.c}.
\end{align}
\end{widetext}

\section{Details of second law analysis} 
\label{app:second-law}

The hydrodynamic constitutive relations must satisfy the second law of thermodynamics. That is, there must exist an entropy current $S^\mu$ whose divergence is locally positive semi-definite, i.e.
\begin{equation}\label{eq:2ndlaw}
    \nabla_\mu S^\mu \equiv \Delta \geq 0,
\end{equation}
for all solutions of the hydrodynamic conservation equations, while keeping the crystal fields $\phi^I$ offshell~\cite{Armas:2019sbe}. We can convert the second law into an off-shell statement by including combinations of the conservation equations 
\begin{align}
    &\nabla_\mu S^\mu + \frac{u_\nu}{T}\lb \nabla_\mu T^{\mu\nu}
    - \tilde F^{\nu\rho} J_\rho
    - F_I^{\nu\rho} J^I_\rho
    - K_I e^{I\nu} \rb \nn\\
    &+ \frac{\mu}{T}\nabla_\mu J^\mu 
    + \frac{\mu_I}{T}\lb \nabla_\mu J^{\mu I} +  e^I_\mu J^\mu \rb
    \equiv \Delta \geq 0,
    \label{eq:offshell-second-law}
\end{align}
where the $K_I e^{I\nu}$ term accounts for the energy-momentum imparted by the offshell configurations of $\phi^I$.
Defining the free energy current
\begin{equation}
    N^\mu = S^\mu + \frac1T \lb T^{\mu\nu}u_\nu
    + \mu J^\mu + \mu_I J^{I\mu} \rb,
\end{equation}
we can massage \cref{eq:offshell-second-law} into the \emph{adiabaticity equation}
\begin{align}\label{eq:adiabaticity}
    \nabla_\mu N^\mu
    &= \half T^{\mu\nu}\delta_\scB g_{\mu\nu}
    + J^\mu \lb \delta_\scB A_\mu 
    + \phi^I \delta_\scB A_{I\mu}
    \rb \nn\\
    &
    \qquad
    + J^{\mu I}\delta_\scB A_{I\mu}
    + K_I \delta_\scB\phi^I
    + \Delta.
\end{align}
Here $\delta_\scB$ denotes a diffeomorphism along $\beta^\mu$, monopole transformation along $\Lambda^\beta$, and a crystal-dipole transformation along $\Lambda_I^\beta$, defined in terms of the hydrodynamic fields using
\begin{align}\label{eq:hydro-variables}
    \beta^\mu 
    &= \frac{u^\mu}{T}, \nn\\
    \Lambda^\beta + \beta^\mu A_\mu 
    &= \frac{\mu}{T}-\phi^I\frac{\mu_I}{T}, \nn\\
    \Lambda^\beta_I + \beta^\mu A_{I\mu} 
    &= \frac{\mu_I}{T},
\end{align}
leading to
\begin{align}
    \delta_\scB g_{\mu\nu} 
    &= 2\nabla_{(\mu}\beta_{\nu)}, \nn\\
    \delta_\scB A_\mu 
    + \phi^I \delta_\scB A_{I\mu}
    &= \dow_\mu \frac{\mu}{T}
    - \frac{\mu_I}{T} e^I_\mu
    + \beta^\lambda \tilde F_{\lambda\mu}, \nn\\
    \delta_\scB A_{I\mu}
    &= \dow_\mu \frac{\mu_I}{T} + \beta^\lambda F_{I\lambda\mu}~, \nn\\
    \delta_\scB\phi^I 
    &= \beta^\mu e^I_\mu.
\end{align}
The constitutive relations in \cref{eq:consti,eq:KI} satisfy the adiabaticity equation with
\begin{align}
    N^\mu &= p\,\beta^\mu, \nn\\
    \Delta &= 
    \frac{\eta}{2T} \lb 2\Delta_\mu^\rho\Delta_\nu^\sigma \nabla_{(\rho}u_{\sigma)} 
    - {\txp\frac{2}{d}}\Delta_{\mu\nu}\nabla_\rho u^\rho\rb^2 
    + \frac{\zeta}{T} \lb \nabla_\mu u^\mu \rb^2 \nn\\
    &~~
    + \frac{\sigma}{T} \lb T\Delta_\mu^\nu\dow_\nu\frac{\mu}{T}
    - \mu_I \Delta_\mu^\nu e^I_\nu - \tilde F_{\mu\rho}u^\rho \rb^2 \nn\\
    &~~
    + \frac{\sigma_{\rm d}}{T}
    \lb T\Delta_\mu^\nu\dow_\nu \frac{\mu_I}{T} - F_{J\mu\rho} u^\rho \rb^2
    + \frac{\sigma_\phi}{T} \lb u^\mu e_\mu^I \rb^2.
\end{align}
Note that the rate of entropy production $\Delta$ is expressed as a quadratic form.
So the production of entropy production is guaranteed, provided that we take the dissipative transport coefficients $\eta$, $\zeta$, $\sigma$, $\sigma_{\rm d}$, and $\sigma_\phi$ to be non-negative. Note that we have only included a select few transport parameters in the constitutive relations in \cref{eq:consti,eq:KI} for simplicity. In general, the adiabaticity equation \eqref{eq:adiabaticity} allows for much more general form of constitutive relations that may be obtained following the construction of \cite{Armas:2019sbe, Armas:2020bmo}. We leave a comprehensive analysis of these for future work.

The procedure needs to be slightly modified for p-wave and s-wave phases, by imposing the local second law \eqref{eq:2ndlaw} for offshell configurations of the respective Goldstones $\varphi_I$ and $\varphi$. For the p-wave phase, we define the operator $K^I_\varphi$ conjugate to $\varphi_I$ using $\delta S \sim \int_{d+1} K_\varphi^I\delta\varphi_I$, which is set to zero when the Goldstone $\varphi_I$ is taken onshell. For offshell configurations of $\varphi_I$, it contributes to energy-momentum and crystal-dipole conservation equations via the source terms $K_\varphi^I\xi_I^\nu$ and $-K_\varphi^I$ respectively. Accounting for these in the offshell second law \eqref{eq:offshell-second-law}, we are led to the p-wave adiabaticity equation
\begin{align}\label{eq:adiabaticity-pwave}
    \nabla_\mu N^\mu
    &= \half T^{\mu\nu}\delta_\scB g_{\mu\nu}
    + J^\mu \lb \delta_\scB A_\mu 
    + \phi^I \delta_\scB A_{I\mu}
    \rb \nn\\
    &~~
    + J^{\mu I}\delta_\scB A_{I\mu}
    + K_I \delta_\scB\phi^I
    + K^I_\varphi \delta_\scB\varphi_I
    + \Delta,
\end{align}
where
\begin{align}
    \delta_\scB\varphi_I
    &= \beta^\mu\xi_{I\mu} - \frac{\mu_I}{T}.
\end{align}
The minimal extension of the constitutive relations in \cref{eq:consti,eq:KI} for the p-wave phase is given as
\begin{align}
    T^{\mu\nu}
    &= \ldots - 2f^{IJ} e^{(\mu}_J\xi_I^{~\nu)}, \nn\\
    J^\mu
    &= \ldots, \nn\\
    J^{\mu I}
    &= \ldots + f^{IJ} e_J^\mu, \nn\\
    K_I 
    &= \ldots - \nabla_\mu \!\lb f^{J}_{~I} \xi_{J}^{~\mu}\rb = 0, \nn\\
    K^I_\varphi 
    &= - \nabla_\mu\!\lb f^{IJ}e_J^\mu\rb 
    - \sigma_\varphi \lb u^\mu \xi^I_\mu - \mu^I \rb = 0,
\end{align}
which leads to an additional contribution to entropy production quadratic form
\begin{align}
    \Delta \sim \frac{\sigma_\varphi}{T} \lb u^\mu \xi_{I\mu} - \mu_I \rb^2,
\end{align}
imposing $\sigma_\varphi$ to be non-negative.

In the s-wave phase, the operator $K_\varphi$ conjugate to $\varphi$ is defined via $\delta S \sim \int_{d+1} K_\varphi(\delta\varphi - \xi_I \delta\phi^I)$, where $\xi_I = \xi_\mu e^\mu_I$. This coupling ensures that $K_I$ remains gauge-invariant. For offshell configurations of $\varphi$, the energy-momentum and charge conservation equations admit source terms $-K_\varphi v^\nu v_\rho\xi^\rho$ and $-K_\varphi$ respectively, and we are led to the s-wave adiabaticity equation
\begin{align}
    \nabla_\mu N^\mu
    &= \half T^{\mu\nu}\delta_\scB g_{\mu\nu}
    + J^\mu \lb \delta_\scB A_\mu 
    + \phi^I \delta_\scB A_{I\mu}
    \rb \nn\\
    &~~
    + J^{I\mu}\delta_\scB A_{I\mu}
    + K_I \delta_\scB \phi^I
    + K_\varphi \lb\delta_\scB \varphi - \xi_I \delta_\scB \phi^I\rb \nn\\
    &~~
    + \Delta,
\end{align}
where
\begin{align}
    \delta_\scB\varphi 
    - \xi_I \delta_\scB \phi^I
    &= - \beta^\nu v_\nu v^\mu \xi_\mu - \frac{\mu}{T}.
\end{align}
In an ordinary superfluid, the above expression would simply be $\beta^\mu \xi_\mu - \mu/T$, which is not crystal-dipole-invariant.
The minimal extension of the constitutive relations in \cref{eq:consti,eq:KI} for the s-wave phase is given as
\begin{align}
    T^{\mu\nu}
    &= \ldots - 2f^{IJ} e^{(\mu}_J\xi_I^{~\nu)}
    + 2g^I e_I^{(\mu} v^{\nu)}  \xi_\rho v^\rho, \nn\\
    J^\mu
    &= \ldots + g^I e_I^\mu, \nn\\
    J^{\mu I}
    &= \ldots + f^{IJ} e_J^\mu, \nn\\
    K_I 
    &= \ldots - \nabla_\mu \!\lb f^{J}_{~I} \xi_{J}^{~\mu}\rb
    + \xi_{IJ}g^J
    + \nabla_\mu\!\lb v^\mu v^\nu \xi_\nu g_I\rb = 0, \nn\\
    K_\varphi 
    &= - \nabla_\mu\!\lb g^I e^\mu_I \rb 
    - \sigma_\varphi \lb v^\nu \xi_\nu + \frac{\mu}{u^\mu v_\mu} \rb = 0,
\end{align}
where $g^I = -\nabla_\mu\!\lb f^{IJ}e_J^\mu\rb$. The additional contribution to entropy production is given as
\begin{align}
    \Delta \sim \frac{\sigma_\varphi}{T} (-u^\mu v_\mu)
    \lb v^\nu \xi_\nu + \frac{\mu}{u^\mu v_\mu} \rb^2,
\end{align}
with non-negative $\sigma_\varphi$, assuming that $u^\mu v_\mu \leq 0$.

\section{Linearised crystal-dipole hydrodynamics and the subdiffusive mode}
\label{app:mode-analysis}

The linearised hydrodynamic equations in the phase with spontaneously unbroken internal symmetries are given as
\begin{align}
    \dow_t\epsilon 
    &= - w\,\dow_i u^i, \nn\\
    \dow_t u^i
    &= - \frac{\dow^i p}{w} 
    + \frac{\eta}{w}\dow_k\dow^k u^i
    + \frac{\zeta+{\txp\frac{d-2}{d}}\eta}{w}\dow^i\dow_k u^k \nn\\
    &\quad 
    - \frac{G}{w}\dow_k\dow^k \delta\phi^i
    - \frac{B+{\txp\frac{d-2}{d}}G}{w}\dow^i\dow_k \delta\phi^k, \nn\\
    \dow_t n
    &= -n\,\dow_i u^i
    + \sigma T \dow_i\dow^i\frac{\mu}{T}
    - \sigma \dow_i\mu^i, \nn\\
    \dow_t\delta\phi^i
    &=
    - u^i 
    + \frac{n}{\sigma_\phi } \mu^i
    + \frac{G}{\sigma_\phi}\dow_k\dow^k \delta\phi^i
    + \frac{B+{\txp\frac{d-2}{d}}G}{\sigma_\phi}\dow^i\dow_k \delta\phi^k, \nn\\
    \dow_t \mu^i
    &= 
    - \frac{\sigma + n^2/\sigma_\phi}{\chi_{\rm d}} \mu^i
    + \frac{\sigma_{\rm d}}{\chi_{\rm d}} \dow_j \dow^j\mu^i
    + \frac{\sigma}{\chi_{\rm d}} T\dow^i\frac{\mu}{T} \nn\\
    &\quad
    - \frac{n}{\chi_{\rm d}}
    \frac{G}{\sigma_\phi}\dow_k\dow^k \delta\phi^i
    - \frac{n}{\chi_{\rm d}}
    \frac{B+{\txp\frac{d-2}{d}}G}{\sigma_\phi}\dow^i\dow_k \delta\phi^k,
    \label{eq:linearised-eqns}
\end{align}
where $w=\epsilon+p$. In deriving these, we have used the form of equation of state
\begin{align}
    p &= p_f(T,\mu)
    - \half \lb B-{\txp\frac{2}{d}}G \rb (\kappa^I_{~I})^2
    - G\, (\kappa_{IJ})^2 \nn\\
    &\qquad
    + \half\chi_{\rm d}(\mu_I)^2
    + \ldots,
\end{align}
where ellipses denote higher order terms in $\kappa_{IJ}$ and $\mu_I$. This results in
\begin{align}
    r_{IJ}
    &= \lb B-{\txp\frac{2}{d}}G \rb \kappa^K_{~K} h_{IJ}
    + 2G\, \kappa_{IJ}
    + \ldots, \nn\\
    n^I 
    &= \chi_{\rm d}\,\mu^I + \ldots.
\end{align}
We can use these to obtain the linearised dispersion relations. Due to the last equation in \cref{eq:linearised-eqns}, the fluctuations of $\mu_i$ are gapped with the frequency scale
\begin{align}
    \frac{\sigma + n^2/\sigma_\phi}{\chi_{\rm d}}.
\end{align}
The remaining four equations give rise to low-energy hydrodynamic modes as discussed in the main text. The details are given in the supplementary Mathematica notebook~\cite{nb-hydro}. Particularly interesting among these is the subdiffusive mode given in \cref{eq:subdiff}. To see the fluctuations that characterise this mode, we can switch to Fourier space and substitute $\omega\sim -iD_{\rm d}k^4$ in \cref{eq:subdiff}. Solving these equations perturbatively in $k$, we find
\begin{widetext}
\begin{align}
    \delta T
    &=
    \frac{T\sigma_\phi D_{\rm d}}{\epsilon+p} \lB
    \frac{1}{\epsilon+p}\frac{\dow\epsilon}{\dow\mu}
    - \frac{n}{\sigma_\phi \sigma} 
    \lb \frac{\dow n}{\dow\mu}
    - \frac{n}{\epsilon+p}\frac{\dow\epsilon}{\dow\mu} \rb
    + \frac{n}{B'} \rB
    k^2\delta\mu, \nn\\
    u_i
    &= 
    \lB -\frac{D_{\rm d}}{\epsilon+p}
    \frac{\dow\epsilon}{\dow\mu} 
    + \cO(k^2)
    \rB ik_ik^2\delta\mu
    , \nn\\
    \delta\kappa_{ij}
    &= \frac{1}{B'}
    \lB n +
    \frac{Ts\sigma_\phi D_{\rm d}}{\epsilon+p} \lb
    \frac{1}{\epsilon+p}\frac{\dow\epsilon}{\dow\mu}
    - \frac{n}{\sigma_\phi \sigma} 
    \lb \frac{\dow n}{\dow\mu}
    - \frac{n}{\epsilon+p}\frac{\dow\epsilon}{\dow\mu} \rb
    + \frac{n}{B'} \rb
    k^2 + \cO(k^4)
    \rB \frac{k_ik_j}{k^2}\delta\mu, \nn\\
     \mu_i
    &=
    \lB 1
    - D_{\rm d} \lb
    \frac{\mu\sigma_\phi}{\epsilon+p}
    \frac{n}{B'} 
    + \frac{\mu\sigma_\phi}{(\epsilon+p)^2}
    \frac{\dow\epsilon}{\dow\mu}
    + \frac{Ts}{\epsilon+p}\frac{1}{\sigma} 
    \lb \frac{\dow n}{\dow\mu}
    - \frac{n}{\epsilon+p}\frac{\dow\epsilon}{\dow\mu} \rb
     \rb k^2
     + \cO(k^4)
     \rB 
     ik_i \delta \mu,
\end{align}
\end{widetext}
together with 
\begin{align}
    D_{\rm d} = \frac{\sigma_{\rm d}}{\dow n/\dow\mu + n^2/B'}.
\end{align}
We see that, at leading order in $k$, this mode is carried by $\delta\mu$ and $\kappa^i_{~i}=n/B'\delta\mu + \ldots$.

\section{Aristotelian geometries from fixed crystals}
\label{app:aristotle}

In this appendix we explain how the Aristotelian geometric structure for coupling to spatial-dipole-invariant field theories in~\cite{Jain:2021ibh, Bidussi:2021nmp} arises from crystal-dipole symmetry when the crystal is fixed. We start by splitting the Lorentz indices $\alpha,\beta,\ldots$ in the tangent space into a time-index $t$ and spatial-indices $a,b,\ldots$. Ordinarily, the Lorentz boost symmetry $\Omega^i_{~t}$ relates the two sectors, but we can explicitly fix this by imposing
\begin{align}
    e^I_\mu \tau^\mu_t = 0,
\end{align}
explicitly breaking the boost symmetry.
We also impose
\begin{align}
    e^I_\mu \tau^\mu_a = \delta^I_a,
\end{align}
which ties the crystal-space rotations $\Omega^I_{~J}$ to the spatial rotations $\Omega^a_{~b}$. In practise, this means that the spatial components of the vielbein $\tau^a_\mu$ are identified with the crystal frame fields $e^I_\mu$. The time component of the vielbein $n_\mu \equiv \tau_\mu^t$ remains independent. Further using the conditions $v^\mu e_\mu^I=0$ and $v^\mu v_\mu =-1$, we can immediately read out $v^\mu\tau_\mu^a = 0$ and $v^\mu n_\mu = 1$. In other words, the collection $(n_\mu,\tau^a_\mu, v^\mu, \tau^\mu_a)$ satisfies
\begin{align}
    n_\mu v^\mu = 1, \quad 
    v^\mu \tau_\mu^a = 0, \quad 
    n_\mu \tau^\mu_a = 0, \quad 
    \tau_\mu^a \tau^\mu_b = \delta^a_b,
\end{align}
and makes up the geometric ingredients of a boost-agnostic Aristotelian spacetime. They transform covariantly under diffeomorphisms $\chi^\mu$ and rotations $\Omega^a_{~b}$. To wit
\begin{align}
    \delta_{\scX} n_\mu 
    &= \lie_\chi n_\mu, \nn\\
    \delta_{\scX} \tau^a_\mu 
    &= \lie_\chi \tau^a_\mu 
    - \Omega^{a}_{~b}\tau^b_\mu,
\end{align}
and similarly for $v^\mu$ and $\tau^\mu_a$. Since we have explicitly fixed the Lorentz boost symmetry, the time components of the spin connection $\omega^t_{~a\mu}$ may consistently be set to zero. The remaining spatial components $\omega^a_{~b\mu}$ transform as
\begin{align}
    \delta_{\scX} \omega^{a}_{~b\mu}
    &= \lie_\chi \omega^{a}_{~b\mu}
    + \dow_\mu \Omega^{a}_{~b}
    - \Omega^{a}_{~c}\omega^{c}_{~b\mu}
    + \omega^{a}_{~c\mu}\Omega^{c}_{~b}.
\end{align}

When $\phi^I$ is non-dynamical, we can replace the monopole gauge field $A_\mu$ with $A'_\mu = A_\mu + \phi^I A_{I\mu}$. Note that $A'_\mu$ would not be a pure background field for dynamical crystals because of its dependence on dynamical $\phi^I$, which is why this procedure can only be implemented for fixed background crystals. $A'_\mu$ transforms as
\begin{align}
    \delta_\scX A'_\mu 
    = \lie_\chi A'_\mu + \dow_\mu \Lambda' + \tau_\mu^a \Lambda'_a ,
\end{align}
where $\Lambda' = \Lambda + \phi^I\Lambda_I$ and $\Lambda'_a = -\delta_a^I\Lambda_I$. In this language, the crystal-dipole gauge field $A_{I\mu}$ becomes the spatial-dipole gauge field $A'_{a\mu}\equiv -\delta_a^I A_{I\mu}$, introduced in~\cite{Jain:2021ibh}. However, to determine the symmetry transformation of $A'_{a\mu}$, we need to first fix the crystal spin-connection $\omega^I_{~J\mu}$. Given that the crystal-space rotations $\Omega^I_{~J}$ have been identified with the spatial rotations $\Omega^a_{~b}$, it is natural to simply identify $\omega^I_{~J\mu}$ with the crystal spin-connection $\omega^a_{~b\mu}$ as well. In this case, $A'_{a\mu}$ transforms as
\begin{align}
    \delta_{\scX} A'_{a\mu}
    &= \lie_\chi A'_{a\mu} 
    + \dow_\mu\Lambda'_a
    - \Lambda'_b \omega^b_{~a\mu} 
    + A'_{b\mu}\Omega^b_{~a}.
\end{align}
This choice comes with a small caveat. Consider the antisymmetric combination 
\begin{align}
    G_{\mu\nu} \equiv F'_{\mu\nu} - 2A'_{a[\mu} \tau^a_{\nu]},
\end{align}
where $F'_{\mu\nu} = 2\dow_{[\mu}A'_{\nu]}$, which transforms homogeneously under dipole transformations as $G_{\mu\nu}\to G_{\mu\nu} + \Lambda'_a T^a_{~\mu\nu}$. If we were to specialise to torsionless geometries where $T^a_{~\mu\nu}$ is vanishing, $G_{\mu\nu}$ would become gauge-invariant and we could consistently set it to zero. This gives rise to a simpler set of background fields where only the symmetric components of the dipole gauge field $a_{ab} = 2A'_{a(\mu}\tau^\mu_{b)}$ are independent, as discussed in~\cite{Jain:2021ibh}. However,
\begin{align}
    K^{ab} \equiv 2T^{(a}_{~\mu\nu} \tau^{b)\nu} v^\mu = 2\lie_v \tau^{(a}_\mu \tau^{b)\mu},
\end{align}
does not contain any components of the connection. So, setting $T^a_{~\mu\nu}$ to zero would require imposing a kinematic constraint on $e^a_\mu$, which is undesirable when coupling field theories to background sources. The best one could do is set the spatial spin-connection to
\begin{align}
    \omega^{a}_{~b\mu}
    &= \tau^{a\nu} \dow_{[\mu} \tau_{b\nu]}
    - \tau^{\nu}_b \dow_{[\mu} \tau^a_{\nu]}
    - \tau^{a\rho} \tau^{\nu}_b \tau_{c\mu} \dow_{[\rho}\tau_{\nu]}^c,
\end{align}
with the minimal torsion $T^a_{~\mu\nu} = -K^{ab}\tau_{b[\mu} n_{\nu]}$. However, this still leaves $G_{\mu\nu}v^\nu$ to be gauge-non-invariant.

Interestingly, we have an alternate choice of crystal spin-connection available to us
\begin{align}
    \delta^a_I\delta^J_b\omega^I_{~\mu J}
    = \omega^a_{~b\mu} 
    - \half T^a_{~\rho\nu} \tau^\nu_b \lb \delta_\mu^\rho + v^\rho n_\mu \rb,
\end{align}
which is still consistent with the symmetry transformation because the second term is covariant under rotations. However, $A'_{a\mu}$ transforms as
\begin{align}
    \delta_{\scX} A'_{a\mu}
    &= \lie_\chi A'_{a\mu} 
    + \dow_\mu\Lambda'_a
    - \Lambda'_b \omega^b_{~a\mu}
    + A'_{b\mu}\Omega^b_{~a} \nn\\
    &\qquad 
    + \half\Lambda'_b T^b_{~\rho\nu}\tau^\nu_a \lb \delta_\mu^\rho + v^\rho n_\mu \rb,
\end{align}
engineered such that $G_{\mu\nu}$ is identically gauge-invariant and can be set to zero even for torsional geometries. For minimal torsion, the additional piece in the transformation reduces to $\half n_\mu K_{ab}\Lambda'^b$, as found in \cite{Jain:2021ibh}.

\section{Crystal-dipole holography}
\label{app:holography}

In this appendix, we give the details of how the holographic model in \cref{eq:holofractons} give rise to a subdiffusive mode in its spectrum of quasinormal modes. For simplicity, we will only look at this model in the probe limit. The bulk equations of motion are given as
\begin{alignat}{2}
    \cE^g_{\sM\sN}
    &\equiv R_{\sM\sN} - \half R g_{\sM\sN} + \Lambda g_{\sM\sN}
    - \kappa\, T_{\sM\sN} &&=0, \nn\\
    \cE^\cA_\sN &\equiv \nabla^\sM\tilde\cF_{\sM\sN}
    &&= 0, \nn\\
    \cE^\cA_{I\sN} &\equiv \nabla^\sM\cF_{I\sM\sN}
    + \frac{g_1}{g_0} \delta_{IJ}e^{J\sM} \tilde\cF_{\sM\sN}
    &&= 0, \nn\\
    \cE^\Phi_I &\equiv
    \nabla_\sM \Big(V'(X) \delta_{IJ} \nabla^\sM\Phi^{J}\Big)
    - \frac{1}{2g_0}\cF_{I\sM\sN}\tilde\cF^{\sM\sN}
    &&= 0,
\end{alignat}
where $T_{\sM\sN}$ is the bulk energy-momentum tensor, which we will not write out explicitly. These equations admit the black brane solution given in \cref{eq:black-brane-solution}.

To find the subdiffusive quasinormal mode, let us work in the probe limit with $\kappa\to 0$. In this limit, the matter fields do not backreact on the geometry. Furthermore, let us only turn on
\begin{align}
    \cA + \phi^I \cA_I
    &= \lb \mu_0 z^{d-1} 
    + \delta\cA_t \rb \df t
    + \delta\cA_x \df x
    , \nn\\
    \cA_I
    &= \delta_I^x \Big( \delta\cA_{xt} \df t 
    + \delta\cA_{xx} \df x \Big), \nn\\
    \Phi^I 
    &= \delta^I_i x^i + \delta^I_x \delta\Phi^x.
\end{align}
One may check that the fluctuations introduced above form a closed set under the bulk equations of motion $\cE^\cA_{z}$, $\cE^\cA_t$, $\cE^\cA_{xz}$, $\cE^\cA_{xt}$, $\cA^\Phi_x$, while the remaining components of the equations of motion are trivial. In particular, one may check that the combinations $\dow_t\cE^\cA_z - z^2\dow_z(\cE^\cA_t/z^2)$ and $\dow_t\cE^\cA_{xz} - z^2\dow_z(\cE^\cA_{xt}/z^2)$ give rise to a decoupled pair of differential equations
\begin{widetext}
\begin{align}\label{eq:hol-eqn}
    \frac{1}{z^{d-2}} \dow_z \Big(z^{d-2}f(z)\dow_z a\Big)
    - \lb \frac{d-2}{z} + 2\dow_z\rb \dow_t a
    + \dow_x^2 a
    &= 
    \lambda\dow_x b, \nn\\
    \frac{1}{z^{d-2}} \dow_z \Big(z^{d-2}f(z)\dow_z b\Big)
    - \lb \frac{d-2}{z} + 2\dow_z\rb \dow_t b
    + \dow_x^2b
    &= 
    \Big(\lambda b - \dow_x a\Big),
\end{align}
\end{widetext}
where $\lambda=g_1/g_0$ and we have exchanged the fluctuations $\delta\cA_t$ and $\delta\cA_{xt}$ for
\begin{alignat}{2}
    a &\equiv \frac{1}{z^{d-2}}\tilde\cF_{zt}
    - (d-1)\mu_0
    &&= \frac{1}{z^{d-2}}\dow_z\delta\cA_{t}, \nn\\
    \lambda b &\equiv \frac{1}{z^{d-2}}\cF_{xzt}
    &&= \frac{1}{z^{d-2}}\dow_z\delta\cA_{xt}.
\end{alignat}
The remaining three independent components of the equations of motion $\cE^\cA_t$, $\cE^\cA_{xt}$, $\cA^\Phi_x$ determine the dynamics of $\delta\cA_x$, $\delta\cA_{xx}$, $\delta\Phi^x$ respectively. 

The first equation in \cref{eq:hol-eqn} for $\lambda\to 0$ reduces the ordinary charged model. We can solve these equations numerically with Dirichlet boundary conditions on $\delta\cA_t$ and $\delta\cA_{xt}$:
\begin{align}
    \delta\cA_{t}(z\to 0) = 0, \qquad 
    \delta\cA_{xt}(z\to 0) = 0,
\end{align}
to find the quasinormal modes given in \cref{fig:dipole-holography}. Since the differential equations only depend on the parameters $z_0$, $\lambda$, and the spatial boundary dimensions $d$, we can repeat this exercise with different choices of parameters to figure out the analytic form of the dispersion relations in \cref{eq:hol-subdiff}. See the supplementary Mathematica notebook~\cite{nb-hol}.

\bibliography{mySpires_ajain,extra}

\begin{thebibliography}{64}%
\makeatletter
\providecommand \@ifxundefined [1]{%
 \@ifx{#1\undefined}
}%
\providecommand \@ifnum [1]{%
 \ifnum #1\expandafter \@firstoftwo
 \else \expandafter \@secondoftwo
 \fi
}%
\providecommand \@ifx [1]{%
 \ifx #1\expandafter \@firstoftwo
 \else \expandafter \@secondoftwo
 \fi
}%
\providecommand \natexlab [1]{#1}%
\providecommand \enquote  [1]{``#1''}%
\providecommand \bibnamefont  [1]{#1}%
\providecommand \bibfnamefont [1]{#1}%
\providecommand \citenamefont [1]{#1}%
\providecommand \href@noop [0]{\@secondoftwo}%
\providecommand \href [0]{\begingroup \@sanitize@url \@href}%
\providecommand \@href[1]{\@@startlink{#1}\@@href}%
\providecommand \@@href[1]{\endgroup#1\@@endlink}%
\providecommand \@sanitize@url [0]{\catcode `\\12\catcode `\$12\catcode
  `\&12\catcode `\#12\catcode `\^12\catcode `\_12\catcode `\%12\relax}%
\providecommand \@@startlink[1]{}%
\providecommand \@@endlink[0]{}%
\providecommand \url  [0]{\begingroup\@sanitize@url \@url }%
\providecommand \@url [1]{\endgroup\@href {#1}{\urlprefix }}%
\providecommand \urlprefix  [0]{URL }%
\providecommand \Eprint [0]{\href }%
\providecommand \doibase [0]{https://doi.org/}%
\providecommand \selectlanguage [0]{\@gobble}%
\providecommand \bibinfo  [0]{\@secondoftwo}%
\providecommand \bibfield  [0]{\@secondoftwo}%
\providecommand \translation [1]{[#1]}%
\providecommand \BibitemOpen [0]{}%
\providecommand \bibitemStop [0]{}%
\providecommand \bibitemNoStop [0]{.\EOS\space}%
\providecommand \EOS [0]{\spacefactor3000\relax}%
\providecommand \BibitemShut  [1]{\csname bibitem#1\endcsname}%
\let\auto@bib@innerbib\@empty
\bibitem [{\citenamefont {Chamon}(2005)}]{Chamon:2004lew}%
  \BibitemOpen
  \bibfield  {author} {\bibinfo {author} {\bibfnamefont {C.}~\bibnamefont
  {Chamon}},\ }\bibfield  {title} {\bibinfo {title} {{Quantum Glassiness}},\
  }\href {https://doi.org/10.1103/physrevlett.94.040402} {\bibfield  {journal}
  {\bibinfo  {journal} {Phys. Rev. Lett.}\ }\textbf {\bibinfo {volume} {94}},\
  \bibinfo {pages} {040402} (\bibinfo {year} {2005})},\ \Eprint
  {https://arxiv.org/abs/cond-mat/0404182} {arXiv:cond-mat/0404182}
  \BibitemShut {NoStop}%
\bibitem [{\citenamefont {{Bravyi}}\ \emph {et~al.}(2011)\citenamefont
  {{Bravyi}}, \citenamefont {{Leemhuis}},\ and\ \citenamefont
  {{Terhal}}}]{2011AnPhy.326..839B}%
  \BibitemOpen
  \bibfield  {author} {\bibinfo {author} {\bibfnamefont {S.}~\bibnamefont
  {{Bravyi}}}, \bibinfo {author} {\bibfnamefont {B.}~\bibnamefont
  {{Leemhuis}}},\ and\ \bibinfo {author} {\bibfnamefont {B.~M.}\ \bibnamefont
  {{Terhal}}},\ }\bibfield  {title} {\bibinfo {title} {{Topological order in an
  exactly solvable 3D spin model}},\ }\href
  {https://doi.org/10.1016/j.aop.2010.11.002} {\bibfield  {journal} {\bibinfo
  {journal} {Annals of Physics}\ }\textbf {\bibinfo {volume} {326}},\ \bibinfo
  {pages} {839} (\bibinfo {year} {2011})},\ \Eprint
  {https://arxiv.org/abs/1006.4871} {arXiv:1006.4871 [quant-ph]} \BibitemShut
  {NoStop}%
\bibitem [{\citenamefont {Haah}(2011)}]{Haah:2011drr}%
  \BibitemOpen
  \bibfield  {author} {\bibinfo {author} {\bibfnamefont {J.}~\bibnamefont
  {Haah}},\ }\bibfield  {title} {\bibinfo {title} {{Local stabilizer codes in
  three dimensions without string logical operators}},\ }\href
  {https://doi.org/10.1103/physreva.83.042330} {\bibfield  {journal} {\bibinfo
  {journal} {Phys. Rev. A}\ }\textbf {\bibinfo {volume} {83}},\ \bibinfo
  {pages} {042330} (\bibinfo {year} {2011})},\ \Eprint
  {https://arxiv.org/abs/1101.1962} {arXiv:1101.1962 [quant-ph]} \BibitemShut
  {NoStop}%
\bibitem [{\citenamefont {Vijay}\ \emph {et~al.}(2015)\citenamefont {Vijay},
  \citenamefont {Haah},\ and\ \citenamefont {Fu}}]{Vijay:2015mka}%
  \BibitemOpen
  \bibfield  {author} {\bibinfo {author} {\bibfnamefont {S.}~\bibnamefont
  {Vijay}}, \bibinfo {author} {\bibfnamefont {J.}~\bibnamefont {Haah}},\ and\
  \bibinfo {author} {\bibfnamefont {L.}~\bibnamefont {Fu}},\ }\bibfield
  {title} {\bibinfo {title} {{A New Kind of Topological Quantum Order: A
  Dimensional Hierarchy of Quasiparticles Built from Stationary Excitations}},\
  }\href {https://doi.org/10.1103/PhysRevB.92.235136} {\bibfield  {journal}
  {\bibinfo  {journal} {Phys. Rev. B}\ }\textbf {\bibinfo {volume} {92}},\
  \bibinfo {pages} {235136} (\bibinfo {year} {2015})},\ \Eprint
  {https://arxiv.org/abs/1505.02576} {arXiv:1505.02576 [cond-mat.str-el]}
  \BibitemShut {NoStop}%
\bibitem [{\citenamefont {Nandkishore}\ and\ \citenamefont
  {Hermele}(2019)}]{Nandkishore:2018sel}%
  \BibitemOpen
  \bibfield  {author} {\bibinfo {author} {\bibfnamefont {R.~M.}\ \bibnamefont
  {Nandkishore}}\ and\ \bibinfo {author} {\bibfnamefont {M.}~\bibnamefont
  {Hermele}},\ }\bibfield  {title} {\bibinfo {title} {{Fractons}},\ }\href
  {https://doi.org/10.1146/annurev-conmatphys-031218-013604} {\bibfield
  {journal} {\bibinfo  {journal} {Ann. Rev. Condensed Matter Phys.}\ }\textbf
  {\bibinfo {volume} {10}},\ \bibinfo {pages} {295} (\bibinfo {year} {2019})},\
  \Eprint {https://arxiv.org/abs/1803.11196} {arXiv:1803.11196
  [cond-mat.str-el]} \BibitemShut {NoStop}%
\bibitem [{\citenamefont {Pretko}\ \emph {et~al.}(2020)\citenamefont {Pretko},
  \citenamefont {Chen},\ and\ \citenamefont {You}}]{Pretko:2020cko}%
  \BibitemOpen
  \bibfield  {author} {\bibinfo {author} {\bibfnamefont {M.}~\bibnamefont
  {Pretko}}, \bibinfo {author} {\bibfnamefont {X.}~\bibnamefont {Chen}},\ and\
  \bibinfo {author} {\bibfnamefont {Y.}~\bibnamefont {You}},\ }\bibfield
  {title} {\bibinfo {title} {{Fracton Phases of Matter}},\ }\bibfield
  {journal} {\bibinfo  {journal} {preprint}\ }\href
  {https://doi.org/10.1142/S0217751X20300033} {10.1142/S0217751X20300033}
  (\bibinfo {year} {2020}),\ \Eprint {https://arxiv.org/abs/2001.01722}
  {arXiv:2001.01722 [cond-mat.str-el]} \BibitemShut {NoStop}%
\bibitem [{\citenamefont {Vijay}\ \emph {et~al.}(2016)\citenamefont {Vijay},
  \citenamefont {Haah},\ and\ \citenamefont {Fu}}]{Vijay:2016phm}%
  \BibitemOpen
  \bibfield  {author} {\bibinfo {author} {\bibfnamefont {S.}~\bibnamefont
  {Vijay}}, \bibinfo {author} {\bibfnamefont {J.}~\bibnamefont {Haah}},\ and\
  \bibinfo {author} {\bibfnamefont {L.}~\bibnamefont {Fu}},\ }\bibfield
  {title} {\bibinfo {title} {{Fracton Topological Order, Generalized Lattice
  Gauge Theory and Duality}},\ }\href
  {https://doi.org/10.1103/PhysRevB.94.235157} {\bibfield  {journal} {\bibinfo
  {journal} {Phys. Rev. B}\ }\textbf {\bibinfo {volume} {94}},\ \bibinfo
  {pages} {235157} (\bibinfo {year} {2016})},\ \Eprint
  {https://arxiv.org/abs/1603.04442} {arXiv:1603.04442 [cond-mat.str-el]}
  \BibitemShut {NoStop}%
\bibitem [{\citenamefont {You}\ \emph {et~al.}(2018)\citenamefont {You},
  \citenamefont {Devakul}, \citenamefont {Burnell},\ and\ \citenamefont
  {Sondhi}}]{You:2018oai}%
  \BibitemOpen
  \bibfield  {author} {\bibinfo {author} {\bibfnamefont {Y.}~\bibnamefont
  {You}}, \bibinfo {author} {\bibfnamefont {T.}~\bibnamefont {Devakul}},
  \bibinfo {author} {\bibfnamefont {F.~J.}\ \bibnamefont {Burnell}},\ and\
  \bibinfo {author} {\bibfnamefont {S.~L.}\ \bibnamefont {Sondhi}},\ }\bibfield
   {title} {\bibinfo {title} {{Subsystem symmetry protected topological
  order}},\ }\href {https://doi.org/10.1103/PhysRevB.98.035112} {\bibfield
  {journal} {\bibinfo  {journal} {Phys. Rev. B}\ }\textbf {\bibinfo {volume}
  {98}},\ \bibinfo {pages} {035112} (\bibinfo {year} {2018})},\ \Eprint
  {https://arxiv.org/abs/1803.02369} {arXiv:1803.02369 [cond-mat.str-el]}
  \BibitemShut {NoStop}%
\bibitem [{\citenamefont {Seiberg}\ and\ \citenamefont
  {Shao}(2021)}]{Seiberg:2020bhn}%
  \BibitemOpen
  \bibfield  {author} {\bibinfo {author} {\bibfnamefont {N.}~\bibnamefont
  {Seiberg}}\ and\ \bibinfo {author} {\bibfnamefont {S.-H.}\ \bibnamefont
  {Shao}},\ }\bibfield  {title} {\bibinfo {title} {{Exotic Symmetries, Duality,
  and Fractons in 2+1-Dimensional Quantum Field Theory}},\ }\href
  {https://doi.org/10.21468/SciPostPhys.10.2.027} {\bibfield  {journal}
  {\bibinfo  {journal} {SciPost Phys.}\ }\textbf {\bibinfo {volume} {10}},\
  \bibinfo {pages} {027} (\bibinfo {year} {2021})},\ \Eprint
  {https://arxiv.org/abs/2003.10466} {arXiv:2003.10466 [cond-mat.str-el]}
  \BibitemShut {NoStop}%
\bibitem [{\citenamefont {Seiberg}\ and\ \citenamefont
  {Shao}(2020)}]{Seiberg:2020wsg}%
  \BibitemOpen
  \bibfield  {author} {\bibinfo {author} {\bibfnamefont {N.}~\bibnamefont
  {Seiberg}}\ and\ \bibinfo {author} {\bibfnamefont {S.-H.}\ \bibnamefont
  {Shao}},\ }\bibfield  {title} {\bibinfo {title} {{Exotic $U(1)$ Symmetries,
  Duality, and Fractons in 3+1-Dimensional Quantum Field Theory}},\ }\href
  {https://doi.org/10.21468/SciPostPhys.9.4.046} {\bibfield  {journal}
  {\bibinfo  {journal} {SciPost Phys.}\ }\textbf {\bibinfo {volume} {9}},\
  \bibinfo {pages} {046} (\bibinfo {year} {2020})},\ \Eprint
  {https://arxiv.org/abs/2004.00015} {arXiv:2004.00015 [cond-mat.str-el]}
  \BibitemShut {NoStop}%
\bibitem [{\citenamefont {Pretko}(2017)}]{Pretko:2016kxt}%
  \BibitemOpen
  \bibfield  {author} {\bibinfo {author} {\bibfnamefont {M.}~\bibnamefont
  {Pretko}},\ }\bibfield  {title} {\bibinfo {title} {{Subdimensional Particle
  Structure of Higher Rank U(1) Spin Liquids}},\ }\href
  {https://doi.org/10.1103/PhysRevB.95.115139} {\bibfield  {journal} {\bibinfo
  {journal} {Phys. Rev. B}\ }\textbf {\bibinfo {volume} {95}},\ \bibinfo
  {pages} {115139} (\bibinfo {year} {2017})},\ \Eprint
  {https://arxiv.org/abs/1604.05329} {arXiv:1604.05329 [cond-mat.str-el]}
  \BibitemShut {NoStop}%
\bibitem [{\citenamefont {Gromov}(2019{\natexlab{a}})}]{Gromov:2018nbv}%
  \BibitemOpen
  \bibfield  {author} {\bibinfo {author} {\bibfnamefont {A.}~\bibnamefont
  {Gromov}},\ }\bibfield  {title} {\bibinfo {title} {{Towards classification of
  Fracton phases: the multipole algebra}},\ }\href
  {https://doi.org/10.1103/PhysRevX.9.031035} {\bibfield  {journal} {\bibinfo
  {journal} {Phys.\ Rev.\ X}\ }\textbf {\bibinfo {volume} {9}},\ \bibinfo
  {pages} {031035} (\bibinfo {year} {2019}{\natexlab{a}})},\ \Eprint
  {https://arxiv.org/abs/1812.05104} {arXiv:1812.05104 [cond-mat.str-el]}
  \BibitemShut {NoStop}%
\bibitem [{\citenamefont {Slagle}\ \emph {et~al.}(2019)\citenamefont {Slagle},
  \citenamefont {Prem},\ and\ \citenamefont {Pretko}}]{Slagle:2018kqf}%
  \BibitemOpen
  \bibfield  {author} {\bibinfo {author} {\bibfnamefont {K.}~\bibnamefont
  {Slagle}}, \bibinfo {author} {\bibfnamefont {A.}~\bibnamefont {Prem}},\ and\
  \bibinfo {author} {\bibfnamefont {M.}~\bibnamefont {Pretko}},\ }\bibfield
  {title} {\bibinfo {title} {{Symmetric Tensor Gauge Theories on Curved
  Spaces}},\ }\href {https://doi.org/10.1016/j.aop.2019.167910} {\bibfield
  {journal} {\bibinfo  {journal} {Annals Phys.}\ }\textbf {\bibinfo {volume}
  {410}},\ \bibinfo {pages} {167910} (\bibinfo {year} {2019})},\ \Eprint
  {https://arxiv.org/abs/1807.00827} {arXiv:1807.00827 [cond-mat.str-el]}
  \BibitemShut {NoStop}%
\bibitem [{\citenamefont {Jain}\ and\ \citenamefont
  {Jensen}(2022)}]{Jain:2021ibh}%
  \BibitemOpen
  \bibfield  {author} {\bibinfo {author} {\bibfnamefont {A.}~\bibnamefont
  {Jain}}\ and\ \bibinfo {author} {\bibfnamefont {K.}~\bibnamefont {Jensen}},\
  }\bibfield  {title} {\bibinfo {title} {{Fractons in curved space}},\ }\href
  {https://doi.org/10.21468/SciPostPhys.12.4.142} {\bibfield  {journal}
  {\bibinfo  {journal} {SciPost Phys.}\ }\textbf {\bibinfo {volume} {12}},\
  \bibinfo {pages} {142} (\bibinfo {year} {2022})},\ \Eprint
  {https://arxiv.org/abs/2111.03973} {arXiv:2111.03973 [hep-th]} \BibitemShut
  {NoStop}%
\bibitem [{\citenamefont {Pe\~na Benitez}(2023)}]{Pena-Benitez:2021ipo}%
  \BibitemOpen
  \bibfield  {author} {\bibinfo {author} {\bibfnamefont {F.}~\bibnamefont
  {Pe\~na Benitez}},\ }\bibfield  {title} {\bibinfo {title} {{Fractons,
  symmetric gauge fields and geometry}},\ }\href
  {https://doi.org/10.1103/PhysRevResearch.5.013101} {\bibfield  {journal}
  {\bibinfo  {journal} {Phys. Rev. Res.}\ }\textbf {\bibinfo {volume} {5}},\
  \bibinfo {pages} {013101} (\bibinfo {year} {2023})},\ \Eprint
  {https://arxiv.org/abs/2107.13884} {arXiv:2107.13884 [cond-mat.str-el]}
  \BibitemShut {NoStop}%
\bibitem [{\citenamefont {Bidussi}\ \emph {et~al.}(2022)\citenamefont
  {Bidussi}, \citenamefont {Hartong}, \citenamefont {Have}, \citenamefont
  {Musaeus},\ and\ \citenamefont {Prohazka}}]{Bidussi:2021nmp}%
  \BibitemOpen
  \bibfield  {author} {\bibinfo {author} {\bibfnamefont {L.}~\bibnamefont
  {Bidussi}}, \bibinfo {author} {\bibfnamefont {J.}~\bibnamefont {Hartong}},
  \bibinfo {author} {\bibfnamefont {E.}~\bibnamefont {Have}}, \bibinfo {author}
  {\bibfnamefont {J.}~\bibnamefont {Musaeus}},\ and\ \bibinfo {author}
  {\bibfnamefont {S.}~\bibnamefont {Prohazka}},\ }\bibfield  {title} {\bibinfo
  {title} {{Fractons, dipole symmetries and curved spacetime}},\ }\href
  {https://doi.org/10.21468/SciPostPhys.12.6.205} {\bibfield  {journal}
  {\bibinfo  {journal} {SciPost Phys.}\ }\textbf {\bibinfo {volume} {12}},\
  \bibinfo {pages} {205} (\bibinfo {year} {2022})},\ \Eprint
  {https://arxiv.org/abs/2111.03668} {arXiv:2111.03668 [hep-th]} \BibitemShut
  {NoStop}%
\bibitem [{\citenamefont {Glorioso}\ \emph {et~al.}(2023)\citenamefont
  {Glorioso}, \citenamefont {Huang}, \citenamefont {Guo}, \citenamefont
  {Rodriguez-Nieva},\ and\ \citenamefont {Lucas}}]{Glorioso:2023chm}%
  \BibitemOpen
  \bibfield  {author} {\bibinfo {author} {\bibfnamefont {P.}~\bibnamefont
  {Glorioso}}, \bibinfo {author} {\bibfnamefont {X.}~\bibnamefont {Huang}},
  \bibinfo {author} {\bibfnamefont {J.}~\bibnamefont {Guo}}, \bibinfo {author}
  {\bibfnamefont {J.~F.}\ \bibnamefont {Rodriguez-Nieva}},\ and\ \bibinfo
  {author} {\bibfnamefont {A.}~\bibnamefont {Lucas}},\ }\bibfield  {title}
  {\bibinfo {title} {{Goldstone bosons and fluctuating hydrodynamics with
  dipole and momentum conservation}},\ }\href
  {https://doi.org/10.1007/JHEP05(2023)022} {\bibfield  {journal} {\bibinfo
  {journal} {JHEP}\ }\textbf {\bibinfo {volume} {05}}\bibfield  {number}
  {\bibinfo  {number} { (05)},\ \bibinfo {pages} {022}},\ }\Eprint
  {https://arxiv.org/abs/2301.02680} {arXiv:2301.02680 [hep-th]} \BibitemShut
  {NoStop}%
\bibitem [{\citenamefont {Jensen}\ and\ \citenamefont
  {Raz}(2024)}]{Jensen:2022ibn}%
  \BibitemOpen
  \bibfield  {author} {\bibinfo {author} {\bibfnamefont {K.}~\bibnamefont
  {Jensen}}\ and\ \bibinfo {author} {\bibfnamefont {A.}~\bibnamefont {Raz}},\
  }\bibfield  {title} {\bibinfo {title} {{Large N Fractons}},\ }\href
  {https://doi.org/10.1103/PhysRevLett.132.071603} {\bibfield  {journal}
  {\bibinfo  {journal} {Phys. Rev. Lett.}\ }\textbf {\bibinfo {volume} {132}},\
  \bibinfo {pages} {071603} (\bibinfo {year} {2024})},\ \Eprint
  {https://arxiv.org/abs/2205.01132} {arXiv:2205.01132 [hep-th]} \BibitemShut
  {NoStop}%
\bibitem [{\citenamefont {Gromov}\ \emph {et~al.}(2020)\citenamefont {Gromov},
  \citenamefont {Lucas},\ and\ \citenamefont {Nandkishore}}]{Gromov:2020yoc}%
  \BibitemOpen
  \bibfield  {author} {\bibinfo {author} {\bibfnamefont {A.}~\bibnamefont
  {Gromov}}, \bibinfo {author} {\bibfnamefont {A.}~\bibnamefont {Lucas}},\ and\
  \bibinfo {author} {\bibfnamefont {R.~M.}\ \bibnamefont {Nandkishore}},\
  }\bibfield  {title} {\bibinfo {title} {{Fracton hydrodynamics}},\ }\href
  {https://doi.org/10.1103/PhysRevResearch.2.033124} {\bibfield  {journal}
  {\bibinfo  {journal} {Phys. Rev. Res.}\ }\textbf {\bibinfo {volume} {2}},\
  \bibinfo {pages} {033124} (\bibinfo {year} {2020})},\ \Eprint
  {https://arxiv.org/abs/2003.09429} {arXiv:2003.09429 [cond-mat.str-el]}
  \BibitemShut {NoStop}%
\bibitem [{\citenamefont {Osborne}\ and\ \citenamefont
  {Lucas}(2022)}]{Osborne:2021mej}%
  \BibitemOpen
  \bibfield  {author} {\bibinfo {author} {\bibfnamefont {A.}~\bibnamefont
  {Osborne}}\ and\ \bibinfo {author} {\bibfnamefont {A.}~\bibnamefont
  {Lucas}},\ }\bibfield  {title} {\bibinfo {title} {{Infinite families of
  fracton fluids with momentum conservation}},\ }\href
  {https://doi.org/10.1103/PhysRevB.105.024311} {\bibfield  {journal} {\bibinfo
   {journal} {Phys. Rev. B}\ }\textbf {\bibinfo {volume} {105}},\ \bibinfo
  {pages} {024311} (\bibinfo {year} {2022})},\ \Eprint
  {https://arxiv.org/abs/2111.09323} {arXiv:2111.09323 [cond-mat.stat-mech]}
  \BibitemShut {NoStop}%
\bibitem [{\citenamefont {Grosvenor}\ \emph {et~al.}(2021)\citenamefont
  {Grosvenor}, \citenamefont {Hoyos}, \citenamefont {Pe\~na Benitez},\ and\
  \citenamefont {Sur\'owka}}]{Grosvenor:2021rrt}%
  \BibitemOpen
  \bibfield  {author} {\bibinfo {author} {\bibfnamefont {K.~T.}\ \bibnamefont
  {Grosvenor}}, \bibinfo {author} {\bibfnamefont {C.}~\bibnamefont {Hoyos}},
  \bibinfo {author} {\bibfnamefont {F.}~\bibnamefont {Pe\~na Benitez}},\ and\
  \bibinfo {author} {\bibfnamefont {P.}~\bibnamefont {Sur\'owka}},\ }\bibfield
  {title} {\bibinfo {title} {{Hydrodynamics of ideal fracton fluids}},\ }\href
  {https://doi.org/10.1103/PhysRevResearch.3.043186} {\bibfield  {journal}
  {\bibinfo  {journal} {Phys. Rev. Res.}\ }\textbf {\bibinfo {volume} {3}},\
  \bibinfo {pages} {043186} (\bibinfo {year} {2021})},\ \Eprint
  {https://arxiv.org/abs/2105.01084} {arXiv:2105.01084 [cond-mat.str-el]}
  \BibitemShut {NoStop}%
\bibitem [{\citenamefont {Jain}\ \emph {et~al.}(2023)\citenamefont {Jain},
  \citenamefont {Jensen}, \citenamefont {Liu},\ and\ \citenamefont
  {Mefford}}]{Jain:2023nbf}%
  \BibitemOpen
  \bibfield  {author} {\bibinfo {author} {\bibfnamefont {A.}~\bibnamefont
  {Jain}}, \bibinfo {author} {\bibfnamefont {K.}~\bibnamefont {Jensen}},
  \bibinfo {author} {\bibfnamefont {R.}~\bibnamefont {Liu}},\ and\ \bibinfo
  {author} {\bibfnamefont {E.}~\bibnamefont {Mefford}},\ }\bibfield  {title}
  {\bibinfo {title} {{Dipole superfluid hydrodynamics}},\ }\href
  {https://doi.org/10.1007/JHEP09(2023)184} {\bibfield  {journal} {\bibinfo
  {journal} {JHEP}\ }\textbf {\bibinfo {volume} {09}}\bibfield  {number}
  {\bibinfo  {number} { (09)},\ \bibinfo {pages} {184}},\ }\Eprint
  {https://arxiv.org/abs/2304.09852} {arXiv:2304.09852 [hep-th]} \BibitemShut
  {NoStop}%
\bibitem [{\citenamefont {Armas}\ and\ \citenamefont
  {Have}(2024)}]{Armas:2023ouk}%
  \BibitemOpen
  \bibfield  {author} {\bibinfo {author} {\bibfnamefont {J.}~\bibnamefont
  {Armas}}\ and\ \bibinfo {author} {\bibfnamefont {E.}~\bibnamefont {Have}},\
  }\bibfield  {title} {\bibinfo {title} {{Ideal fracton superfluids}},\ }\href
  {https://doi.org/10.21468/SciPostPhys.16.1.039} {\bibfield  {journal}
  {\bibinfo  {journal} {SciPost Phys.}\ }\textbf {\bibinfo {volume} {16}},\
  \bibinfo {pages} {039} (\bibinfo {year} {2024})},\ \Eprint
  {https://arxiv.org/abs/2304.09596} {arXiv:2304.09596 [hep-th]} \BibitemShut
  {NoStop}%
\bibitem [{\citenamefont {Jain}\ \emph {et~al.}(2024)\citenamefont {Jain},
  \citenamefont {Jensen}, \citenamefont {Liu},\ and\ \citenamefont
  {Mefford}}]{Jain:2024kri}%
  \BibitemOpen
  \bibfield  {author} {\bibinfo {author} {\bibfnamefont {A.}~\bibnamefont
  {Jain}}, \bibinfo {author} {\bibfnamefont {K.}~\bibnamefont {Jensen}},
  \bibinfo {author} {\bibfnamefont {R.}~\bibnamefont {Liu}},\ and\ \bibinfo
  {author} {\bibfnamefont {E.}~\bibnamefont {Mefford}},\ }\bibfield  {title}
  {\bibinfo {title} {{Dipole superfluid hydrodynamics. Part II.}},\ }\href
  {https://doi.org/10.1007/JHEP07(2024)197} {\bibfield  {journal} {\bibinfo
  {journal} {JHEP}\ }\textbf {\bibinfo {volume} {07}}\bibfield  {number}
  {\bibinfo  {number} { (07)},\ \bibinfo {pages} {197}},\ }\Eprint
  {https://arxiv.org/abs/2401.16385} {arXiv:2401.16385 [hep-th]} \BibitemShut
  {NoStop}%
\bibitem [{\citenamefont {Pretko}\ and\ \citenamefont
  {Radzihovsky}(2018)}]{Pretko:2018qru}%
  \BibitemOpen
  \bibfield  {author} {\bibinfo {author} {\bibfnamefont {M.}~\bibnamefont
  {Pretko}}\ and\ \bibinfo {author} {\bibfnamefont {L.}~\bibnamefont
  {Radzihovsky}},\ }\bibfield  {title} {\bibinfo {title} {{Fracton-Elasticity
  Duality}},\ }\href {https://doi.org/10.1103/PhysRevLett.120.195301}
  {\bibfield  {journal} {\bibinfo  {journal} {Phys.\ Rev.\ Lett.}\ }\textbf
  {\bibinfo {volume} {120}},\ \bibinfo {pages} {195301} (\bibinfo {year}
  {2018})},\ \Eprint {https://arxiv.org/abs/1711.11044} {arXiv:1711.11044
  [cond-mat.str-el]} \BibitemShut {NoStop}%
\bibitem [{\citenamefont {Nguyen}\ \emph {et~al.}(2020)\citenamefont {Nguyen},
  \citenamefont {Gromov},\ and\ \citenamefont {Moroz}}]{Nguyen:2020yve}%
  \BibitemOpen
  \bibfield  {author} {\bibinfo {author} {\bibfnamefont {D.~X.}\ \bibnamefont
  {Nguyen}}, \bibinfo {author} {\bibfnamefont {A.}~\bibnamefont {Gromov}},\
  and\ \bibinfo {author} {\bibfnamefont {S.}~\bibnamefont {Moroz}},\ }\bibfield
   {title} {\bibinfo {title} {{Fracton-elasticity duality of two-dimensional
  superfluid vortex crystals: defect interactions and quantum melting}},\
  }\href {https://doi.org/10.21468/SciPostPhys.9.5.076} {\bibfield  {journal}
  {\bibinfo  {journal} {SciPost Phys.}\ }\textbf {\bibinfo {volume} {9}},\
  \bibinfo {pages} {076} (\bibinfo {year} {2020})},\ \Eprint
  {https://arxiv.org/abs/2005.12317} {arXiv:2005.12317 [cond-mat.quant-gas]}
  \BibitemShut {NoStop}%
\bibitem [{\citenamefont {Gromov}(2019{\natexlab{b}})}]{Gromov:2017vir}%
  \BibitemOpen
  \bibfield  {author} {\bibinfo {author} {\bibfnamefont {A.}~\bibnamefont
  {Gromov}},\ }\bibfield  {title} {\bibinfo {title} {{Chiral Topological
  Elasticity and Fracton Order}},\ }\href
  {https://doi.org/10.1103/PhysRevLett.122.076403} {\bibfield  {journal}
  {\bibinfo  {journal} {Phys. Rev. Lett.}\ }\textbf {\bibinfo {volume} {122}},\
  \bibinfo {pages} {076403} (\bibinfo {year} {2019}{\natexlab{b}})},\ \Eprint
  {https://arxiv.org/abs/1712.06600} {arXiv:1712.06600 [cond-mat.str-el]}
  \BibitemShut {NoStop}%
\bibitem [{\citenamefont {Caddeo}\ \emph {et~al.}(2022)\citenamefont {Caddeo},
  \citenamefont {Hoyos},\ and\ \citenamefont {Musso}}]{Caddeo:2022ibe}%
  \BibitemOpen
  \bibfield  {author} {\bibinfo {author} {\bibfnamefont {A.}~\bibnamefont
  {Caddeo}}, \bibinfo {author} {\bibfnamefont {C.}~\bibnamefont {Hoyos}},\ and\
  \bibinfo {author} {\bibfnamefont {D.}~\bibnamefont {Musso}},\ }\bibfield
  {title} {\bibinfo {title} {{Emergent dipole gauge fields and fractons}},\
  }\href {https://doi.org/10.1103/PhysRevD.106.L111903} {\bibfield  {journal}
  {\bibinfo  {journal} {Phys. Rev. D}\ }\textbf {\bibinfo {volume} {106}},\
  \bibinfo {pages} {L111903} (\bibinfo {year} {2022})},\ \Eprint
  {https://arxiv.org/abs/2206.12877} {arXiv:2206.12877 [cond-mat.str-el]}
  \BibitemShut {NoStop}%
\bibitem [{\citenamefont {Doshi}\ and\ \citenamefont
  {Gromov}(2020)}]{Doshi:2020jso}%
  \BibitemOpen
  \bibfield  {author} {\bibinfo {author} {\bibfnamefont {D.}~\bibnamefont
  {Doshi}}\ and\ \bibinfo {author} {\bibfnamefont {A.}~\bibnamefont {Gromov}},\
  }\bibfield  {title} {\bibinfo {title} {{Vortices and Fractons}},\ }\href@noop
  {} {\bibfield  {journal} {\bibinfo  {journal} {preprint}\ } (\bibinfo {year}
  {2020})},\ \Eprint {https://arxiv.org/abs/2005.03015} {arXiv:2005.03015
  [cond-mat.str-el]} \BibitemShut {NoStop}%
\bibitem [{\citenamefont {You}\ and\ \citenamefont {von
  Oppen}(2019)}]{You:2018bmf}%
  \BibitemOpen
  \bibfield  {author} {\bibinfo {author} {\bibfnamefont {Y.}~\bibnamefont
  {You}}\ and\ \bibinfo {author} {\bibfnamefont {F.}~\bibnamefont {von
  Oppen}},\ }\bibfield  {title} {\bibinfo {title} {{Majorana Quantum Lego, a
  Route Towards Fracton Matter}},\ }\href
  {https://doi.org/10.1103/PhysRevResearch.1.013011} {\bibfield  {journal}
  {\bibinfo  {journal} {Phys. Rev. Research.}\ }\textbf {\bibinfo {volume}
  {1}},\ \bibinfo {pages} {013011} (\bibinfo {year} {2019})},\ \Eprint
  {https://arxiv.org/abs/1812.06091} {arXiv:1812.06091 [cond-mat.str-el]}
  \BibitemShut {NoStop}%
\bibitem [{\citenamefont {You}\ \emph {et~al.}(2020)\citenamefont {You},
  \citenamefont {Bi},\ and\ \citenamefont {Pretko}}]{You:2019cvs}%
  \BibitemOpen
  \bibfield  {author} {\bibinfo {author} {\bibfnamefont {Y.}~\bibnamefont
  {You}}, \bibinfo {author} {\bibfnamefont {Z.}~\bibnamefont {Bi}},\ and\
  \bibinfo {author} {\bibfnamefont {M.}~\bibnamefont {Pretko}},\ }\bibfield
  {title} {\bibinfo {title} {{Emergent fractons and algebraic quantum liquid
  from plaquette melting transitions}},\ }\href
  {https://doi.org/10.1103/PhysRevResearch.2.013162} {\bibfield  {journal}
  {\bibinfo  {journal} {Phys. Rev. Res.}\ }\textbf {\bibinfo {volume} {2}},\
  \bibinfo {pages} {013162} (\bibinfo {year} {2020})},\ \Eprint
  {https://arxiv.org/abs/1908.08540} {arXiv:1908.08540 [cond-mat.str-el]}
  \BibitemShut {NoStop}%
\bibitem [{\citenamefont {Sous}\ and\ \citenamefont
  {Pretko}(2020)}]{Sous:2019jtt}%
  \BibitemOpen
  \bibfield  {author} {\bibinfo {author} {\bibfnamefont {J.}~\bibnamefont
  {Sous}}\ and\ \bibinfo {author} {\bibfnamefont {M.}~\bibnamefont {Pretko}},\
  }\bibfield  {title} {\bibinfo {title} {{Fractons from polarons}},\ }\href
  {https://doi.org/10.1103/PhysRevB.102.214437} {\bibfield  {journal} {\bibinfo
   {journal} {Phys. Rev. B}\ }\textbf {\bibinfo {volume} {102}},\ \bibinfo
  {pages} {214437} (\bibinfo {year} {2020})},\ \Eprint
  {https://arxiv.org/abs/1904.08424} {arXiv:1904.08424 [cond-mat.str-el]}
  \BibitemShut {NoStop}%
\bibitem [{\citenamefont {Guardado-Sanchez}\ \emph {et~al.}(2020)\citenamefont
  {Guardado-Sanchez}, \citenamefont {Morningstar}, \citenamefont {Spar},
  \citenamefont {Brown}, \citenamefont {Huse},\ and\ \citenamefont
  {Bakr}}]{Guardado-Sanchez:2019bjm}%
  \BibitemOpen
  \bibfield  {author} {\bibinfo {author} {\bibfnamefont {E.}~\bibnamefont
  {Guardado-Sanchez}}, \bibinfo {author} {\bibfnamefont {A.}~\bibnamefont
  {Morningstar}}, \bibinfo {author} {\bibfnamefont {B.~M.}\ \bibnamefont
  {Spar}}, \bibinfo {author} {\bibfnamefont {P.~T.}\ \bibnamefont {Brown}},
  \bibinfo {author} {\bibfnamefont {D.~A.}\ \bibnamefont {Huse}},\ and\
  \bibinfo {author} {\bibfnamefont {W.~S.}\ \bibnamefont {Bakr}},\ }\bibfield
  {title} {\bibinfo {title} {{Subdiffusion and Heat Transport in a Tilted
  Two-Dimensional Fermi-Hubbard System}},\ }\href
  {https://doi.org/10.1103/PhysRevX.10.011042} {\bibfield  {journal} {\bibinfo
  {journal} {Phys. Rev. X}\ }\textbf {\bibinfo {volume} {10}},\ \bibinfo
  {pages} {011042} (\bibinfo {year} {2020})},\ \Eprint
  {https://arxiv.org/abs/1909.05848} {arXiv:1909.05848 [cond-mat.quant-gas]}
  \BibitemShut {NoStop}%
\bibitem [{\citenamefont {Leutwyler}(1997)}]{Leutwyler:1996er}%
  \BibitemOpen
  \bibfield  {author} {\bibinfo {author} {\bibfnamefont {H.}~\bibnamefont
  {Leutwyler}},\ }\bibfield  {title} {\bibinfo {title} {{Phonons as goldstone
  bosons}},\ }\href@noop {} {\bibfield  {journal} {\bibinfo  {journal} {Helv.\
  Phys.\ Acta}\ }\textbf {\bibinfo {volume} {70}},\ \bibinfo {pages} {275}
  (\bibinfo {year} {1997})},\ \Eprint {https://arxiv.org/abs/hep-ph/9609466}
  {arXiv:hep-ph/9609466} \BibitemShut {NoStop}%
\bibitem [{\citenamefont {Nicolis}\ \emph {et~al.}(2015)\citenamefont
  {Nicolis}, \citenamefont {Penco}, \citenamefont {Piazza},\ and\ \citenamefont
  {Rattazzi}}]{Nicolis:2015sra}%
  \BibitemOpen
  \bibfield  {author} {\bibinfo {author} {\bibfnamefont {A.}~\bibnamefont
  {Nicolis}}, \bibinfo {author} {\bibfnamefont {R.}~\bibnamefont {Penco}},
  \bibinfo {author} {\bibfnamefont {F.}~\bibnamefont {Piazza}},\ and\ \bibinfo
  {author} {\bibfnamefont {R.}~\bibnamefont {Rattazzi}},\ }\bibfield  {title}
  {\bibinfo {title} {{Zoology of condensed matter: Framids, ordinary stuff,
  extra-ordinary stuff}},\ }\href {https://doi.org/10.1007/JHEP06(2015)155}
  {\bibfield  {journal} {\bibinfo  {journal} {JHEP}\ }\textbf {\bibinfo
  {volume} {06}}\bibfield  {number} {\bibinfo  {number} { (06)},\ \bibinfo
  {pages} {155}},\ }\Eprint {https://arxiv.org/abs/1501.03845}
  {arXiv:1501.03845 [hep-th]} \BibitemShut {NoStop}%
\bibitem [{\citenamefont {Armas}\ and\ \citenamefont
  {Jain}(2020{\natexlab{a}})}]{Armas:2019sbe}%
  \BibitemOpen
  \bibfield  {author} {\bibinfo {author} {\bibfnamefont {J.}~\bibnamefont
  {Armas}}\ and\ \bibinfo {author} {\bibfnamefont {A.}~\bibnamefont {Jain}},\
  }\bibfield  {title} {\bibinfo {title} {{Viscoelastic hydrodynamics and
  holography}},\ }\href {https://doi.org/10.1007/JHEP01(2020)126} {\bibfield
  {journal} {\bibinfo  {journal} {JHEP}\ }\textbf {\bibinfo {volume}
  {01}}\bibfield  {number} {\bibinfo  {number} { (01)},\ \bibinfo {pages}
  {126}},\ }\Eprint {https://arxiv.org/abs/1908.01175} {arXiv:1908.01175
  [hep-th]} \BibitemShut {NoStop}%
\bibitem [{\citenamefont {Armas}\ and\ \citenamefont
  {Jain}(2020{\natexlab{b}})}]{Armas:2020bmo}%
  \BibitemOpen
  \bibfield  {author} {\bibinfo {author} {\bibfnamefont {J.}~\bibnamefont
  {Armas}}\ and\ \bibinfo {author} {\bibfnamefont {A.}~\bibnamefont {Jain}},\
  }\bibfield  {title} {\bibinfo {title} {{Hydrodynamics for charge density
  waves and their holographic duals}},\ }\href
  {https://doi.org/10.1103/PhysRevD.101.121901} {\bibfield  {journal} {\bibinfo
   {journal} {Phys. Rev. D}\ }\textbf {\bibinfo {volume} {101}},\ \bibinfo
  {pages} {121901} (\bibinfo {year} {2020}{\natexlab{b}})},\ \Eprint
  {https://arxiv.org/abs/2001.07357} {arXiv:2001.07357 [hep-th]} \BibitemShut
  {NoStop}%
\bibitem [{\citenamefont {Armas}\ \emph
  {et~al.}(2023{\natexlab{a}})\citenamefont {Armas}, \citenamefont {Jain},\
  and\ \citenamefont {Lier}}]{Armas:2021vku}%
  \BibitemOpen
  \bibfield  {author} {\bibinfo {author} {\bibfnamefont {J.}~\bibnamefont
  {Armas}}, \bibinfo {author} {\bibfnamefont {A.}~\bibnamefont {Jain}},\ and\
  \bibinfo {author} {\bibfnamefont {R.}~\bibnamefont {Lier}},\ }\bibfield
  {title} {\bibinfo {title} {{Approximate symmetries, pseudo-Goldstones, and
  the second law of thermodynamics}},\ }\href
  {https://doi.org/10.1103/PhysRevD.108.086011} {\bibfield  {journal} {\bibinfo
   {journal} {Phys. Rev. D}\ }\textbf {\bibinfo {volume} {108}},\ \bibinfo
  {pages} {086011} (\bibinfo {year} {2023}{\natexlab{a}})},\ \Eprint
  {https://arxiv.org/abs/2112.14373} {arXiv:2112.14373 [hep-th]} \BibitemShut
  {NoStop}%
\bibitem [{\citenamefont {Armas}\ \emph
  {et~al.}(2023{\natexlab{b}})\citenamefont {Armas}, \citenamefont {van
  Heumen}, \citenamefont {Jain},\ and\ \citenamefont {Lier}}]{Armas:2022vpf}%
  \BibitemOpen
  \bibfield  {author} {\bibinfo {author} {\bibfnamefont {J.}~\bibnamefont
  {Armas}}, \bibinfo {author} {\bibfnamefont {E.}~\bibnamefont {van Heumen}},
  \bibinfo {author} {\bibfnamefont {A.}~\bibnamefont {Jain}},\ and\ \bibinfo
  {author} {\bibfnamefont {R.}~\bibnamefont {Lier}},\ }\bibfield  {title}
  {\bibinfo {title} {{Hydrodynamics of plastic deformations in electronic
  crystals}},\ }\href {https://doi.org/10.1103/PhysRevB.107.155108} {\bibfield
  {journal} {\bibinfo  {journal} {Phys. Rev. B}\ }\textbf {\bibinfo {volume}
  {107}},\ \bibinfo {pages} {155108} (\bibinfo {year} {2023}{\natexlab{b}})},\
  \Eprint {https://arxiv.org/abs/2211.02117} {arXiv:2211.02117
  [cond-mat.str-el]} \BibitemShut {NoStop}%
\bibitem [{\citenamefont {Fukuma}\ and\ \citenamefont
  {Sakatani}(2011)}]{Fukuma:2011pr}%
  \BibitemOpen
  \bibfield  {author} {\bibinfo {author} {\bibfnamefont {M.}~\bibnamefont
  {Fukuma}}\ and\ \bibinfo {author} {\bibfnamefont {Y.}~\bibnamefont
  {Sakatani}},\ }\bibfield  {title} {\bibinfo {title} {{Relativistic
  viscoelastic fluid mechanics}},\ }\href
  {https://doi.org/10.1103/PhysRevE.84.026316} {\bibfield  {journal} {\bibinfo
  {journal} {Phys.\ Rev.\ E}\ }\textbf {\bibinfo {volume} {84}},\ \bibinfo
  {pages} {026316} (\bibinfo {year} {2011})},\ \Eprint
  {https://arxiv.org/abs/1104.1416} {arXiv:1104.1416 [cond-mat.stat-mech]}
  \BibitemShut {NoStop}%
\bibitem [{\citenamefont {Pretko}(2018)}]{Pretko:2018jbi}%
  \BibitemOpen
  \bibfield  {author} {\bibinfo {author} {\bibfnamefont {M.}~\bibnamefont
  {Pretko}},\ }\bibfield  {title} {\bibinfo {title} {{The Fracton Gauge
  Principle}},\ }\href {https://doi.org/10.1103/PhysRevB.98.115134} {\bibfield
  {journal} {\bibinfo  {journal} {Phys.\ Rev.\ B}\ }\textbf {\bibinfo {volume}
  {98}},\ \bibinfo {pages} {115134} (\bibinfo {year} {2018})},\ \Eprint
  {https://arxiv.org/abs/1807.11479} {arXiv:1807.11479 [cond-mat.str-el]}
  \BibitemShut {NoStop}%
\bibitem [{\citenamefont {{Boninsegni}}\ and\ \citenamefont
  {{Prokof'ev}}(2012)}]{2012RvMP...84..759B}%
  \BibitemOpen
  \bibfield  {author} {\bibinfo {author} {\bibfnamefont {M.}~\bibnamefont
  {{Boninsegni}}}\ and\ \bibinfo {author} {\bibfnamefont {N.~V.}\ \bibnamefont
  {{Prokof'ev}}},\ }\bibfield  {title} {\bibinfo {title} {{Colloquium:
  Supersolids: What and where are they?}},\ }\href
  {https://doi.org/10.1103/RevModPhys.84.759} {\bibfield  {journal} {\bibinfo
  {journal} {Reviews of Modern Physics}\ }\textbf {\bibinfo {volume} {84}},\
  \bibinfo {pages} {759} (\bibinfo {year} {2012})},\ \Eprint
  {https://arxiv.org/abs/1201.2227} {arXiv:1201.2227 [cond-mat.stat-mech]}
  \BibitemShut {NoStop}%
\bibitem [{\citenamefont {Stahl}\ \emph {et~al.}(2023)\citenamefont {Stahl},
  \citenamefont {Qi}, \citenamefont {Glorioso}, \citenamefont {Lucas},\ and\
  \citenamefont {Nandkishore}}]{Stahl:2023prt}%
  \BibitemOpen
  \bibfield  {author} {\bibinfo {author} {\bibfnamefont {C.}~\bibnamefont
  {Stahl}}, \bibinfo {author} {\bibfnamefont {M.}~\bibnamefont {Qi}}, \bibinfo
  {author} {\bibfnamefont {P.}~\bibnamefont {Glorioso}}, \bibinfo {author}
  {\bibfnamefont {A.}~\bibnamefont {Lucas}},\ and\ \bibinfo {author}
  {\bibfnamefont {R.}~\bibnamefont {Nandkishore}},\ }\bibfield  {title}
  {\bibinfo {title} {{Fracton superfluid hydrodynamics}},\ }\href
  {https://doi.org/10.1103/PhysRevB.108.144509} {\bibfield  {journal} {\bibinfo
   {journal} {Phys. Rev. B}\ }\textbf {\bibinfo {volume} {108}},\ \bibinfo
  {pages} {144509} (\bibinfo {year} {2023})},\ \Eprint
  {https://arxiv.org/abs/2303.09573} {arXiv:2303.09573 [cond-mat.stat-mech]}
  \BibitemShut {NoStop}%
\bibitem [{Note1()}]{Note1}%
  \BibitemOpen
  \bibinfo {note} {One may also approach hydrodynamics using Schwinger-Keldysh
  effective actions~\cite {Grozdanov:2013dba, Harder:2015nxa, Crossley:2015evo,
  Haehl:2015uoc, Haehl:2018lcu, Jensen:2017kzi, Glorioso:2018wxw}, useful for
  including stochastic fluctuations in hydrodynamic models.}\BibitemShut
  {Stop}%
\bibitem [{nb-(2024{\natexlab{a}})}]{nb-hydro}%
  \BibitemOpen
  \href@noop {} {\bibinfo {title} {See the {Mathematica} notebook}},\ \bibinfo
  {howpublished}
  {\url{https://github.com/ajainphysics/Mathematica-Notebooks/tree/main/\%5BarXiv\%3A2406.07334\%5D\%20Fractonic\%20Solids/FractonicSolids_Modes.nb}}
  (\bibinfo {year} {2024}{\natexlab{a}}),\ \bibinfo {note} {for details of the
  mode spectrum of various phases of crystal-dipole-invariant relativistic
  hydrodynamics.}\BibitemShut {Stop}%
\bibitem [{Note2()}]{Note2}%
  \BibitemOpen
  \bibinfo {note} {There will still be finite wavevector instabilities that
  usually appear in relativistic hydrodynamics and may be treated using similar
  techniques.}\BibitemShut {Stop}%
\bibitem [{\citenamefont {Gorantla}\ \emph {et~al.}(2021)\citenamefont
  {Gorantla}, \citenamefont {Lam}, \citenamefont {Seiberg},\ and\ \citenamefont
  {Shao}}]{Gorantla:2021bda}%
  \BibitemOpen
  \bibfield  {author} {\bibinfo {author} {\bibfnamefont {P.}~\bibnamefont
  {Gorantla}}, \bibinfo {author} {\bibfnamefont {H.~T.}\ \bibnamefont {Lam}},
  \bibinfo {author} {\bibfnamefont {N.}~\bibnamefont {Seiberg}},\ and\ \bibinfo
  {author} {\bibfnamefont {S.-H.}\ \bibnamefont {Shao}},\ }\bibfield  {title}
  {\bibinfo {title} {{Low-energy limit of some exotic lattice theories and
  UV/IR mixing}},\ }\href {https://doi.org/10.1103/PhysRevB.104.235116}
  {\bibfield  {journal} {\bibinfo  {journal} {Phys. Rev. B}\ }\textbf {\bibinfo
  {volume} {104}},\ \bibinfo {pages} {235116} (\bibinfo {year} {2021})},\
  \Eprint {https://arxiv.org/abs/2108.00020} {arXiv:2108.00020
  [cond-mat.str-el]} \BibitemShut {NoStop}%
\bibitem [{\citenamefont {Ganesan}\ and\ \citenamefont
  {Lucas}(2020)}]{Ganesan:2020wvm}%
  \BibitemOpen
  \bibfield  {author} {\bibinfo {author} {\bibfnamefont {K.}~\bibnamefont
  {Ganesan}}\ and\ \bibinfo {author} {\bibfnamefont {A.}~\bibnamefont
  {Lucas}},\ }\bibfield  {title} {\bibinfo {title} {{Holographic
  subdiffusion}},\ }\href {https://doi.org/10.1007/JHEP12(2020)149} {\bibfield
  {journal} {\bibinfo  {journal} {JHEP}\ }\textbf {\bibinfo {volume}
  {12}}\bibfield  {number} {\bibinfo  {number} { (12)},\ \bibinfo {pages}
  {149}},\ }\Eprint {https://arxiv.org/abs/2008.09638} {arXiv:2008.09638
  [hep-th]} \BibitemShut {NoStop}%
\bibitem [{\citenamefont {Baggioli}\ and\ \citenamefont
  {Pujolas}(2015)}]{Baggioli:2014roa}%
  \BibitemOpen
  \bibfield  {author} {\bibinfo {author} {\bibfnamefont {M.}~\bibnamefont
  {Baggioli}}\ and\ \bibinfo {author} {\bibfnamefont {O.}~\bibnamefont
  {Pujolas}},\ }\bibfield  {title} {\bibinfo {title} {{Electron-Phonon
  Interactions, Metal-Insulator Transitions, and Holographic Massive
  Gravity}},\ }\href {https://doi.org/10.1103/PhysRevLett.114.251602}
  {\bibfield  {journal} {\bibinfo  {journal} {Phys.\ Rev.\ Lett.}\ }\textbf
  {\bibinfo {volume} {114}},\ \bibinfo {pages} {251602} (\bibinfo {year}
  {2015})},\ \Eprint {https://arxiv.org/abs/1411.1003} {arXiv:1411.1003
  [hep-th]} \BibitemShut {NoStop}%
\bibitem [{\citenamefont {Alberte}\ \emph {et~al.}(2016)\citenamefont
  {Alberte}, \citenamefont {Baggioli}, \citenamefont {Khmelnitsky},\ and\
  \citenamefont {Pujolas}}]{Alberte:2015isw}%
  \BibitemOpen
  \bibfield  {author} {\bibinfo {author} {\bibfnamefont {L.}~\bibnamefont
  {Alberte}}, \bibinfo {author} {\bibfnamefont {M.}~\bibnamefont {Baggioli}},
  \bibinfo {author} {\bibfnamefont {A.}~\bibnamefont {Khmelnitsky}},\ and\
  \bibinfo {author} {\bibfnamefont {O.}~\bibnamefont {Pujolas}},\ }\bibfield
  {title} {\bibinfo {title} {{Solid Holography and Massive Gravity}},\ }\href
  {https://doi.org/10.1007/JHEP02(2016)114} {\bibfield  {journal} {\bibinfo
  {journal} {JHEP}\ }\textbf {\bibinfo {volume} {02}}\bibfield  {number}
  {\bibinfo  {number} { (02)},\ \bibinfo {pages} {114}},\ }\Eprint
  {https://arxiv.org/abs/1510.09089} {arXiv:1510.09089 [hep-th]} \BibitemShut
  {NoStop}%
\bibitem [{\citenamefont {Baggioli}\ and\ \citenamefont
  {Gout\'eraux}(2023)}]{Baggioli:2022pyb}%
  \BibitemOpen
  \bibfield  {author} {\bibinfo {author} {\bibfnamefont {M.}~\bibnamefont
  {Baggioli}}\ and\ \bibinfo {author} {\bibfnamefont {B.}~\bibnamefont
  {Gout\'eraux}},\ }\bibfield  {title} {\bibinfo {title} {{Colloquium:
  Hydrodynamics and holography of charge density wave phases}},\ }\href
  {https://doi.org/10.1103/RevModPhys.95.011001} {\bibfield  {journal}
  {\bibinfo  {journal} {Rev. Mod. Phys.}\ }\textbf {\bibinfo {volume} {95}},\
  \bibinfo {pages} {011001} (\bibinfo {year} {2023})},\ \Eprint
  {https://arxiv.org/abs/2203.03298} {arXiv:2203.03298 [hep-th]} \BibitemShut
  {NoStop}%
\bibitem [{\citenamefont {Andrade}\ and\ \citenamefont
  {Withers}(2014)}]{Andrade:2013gsa}%
  \BibitemOpen
  \bibfield  {author} {\bibinfo {author} {\bibfnamefont {T.}~\bibnamefont
  {Andrade}}\ and\ \bibinfo {author} {\bibfnamefont {B.}~\bibnamefont
  {Withers}},\ }\bibfield  {title} {\bibinfo {title} {{A simple holographic
  model of momentum relaxation}},\ }\href
  {https://doi.org/10.1007/JHEP05(2014)101} {\bibfield  {journal} {\bibinfo
  {journal} {JHEP}\ }\textbf {\bibinfo {volume} {05}}\bibfield  {number}
  {\bibinfo  {number} { (05)},\ \bibinfo {pages} {101}},\ }\Eprint
  {https://arxiv.org/abs/1311.5157} {arXiv:1311.5157 [hep-th]} \BibitemShut
  {NoStop}%
\bibitem [{\citenamefont {Bhattacharya}\ \emph {et~al.}(2011)\citenamefont
  {Bhattacharya}, \citenamefont {Bhattacharyya},\ and\ \citenamefont
  {Minwalla}}]{Bhattacharya:2011eea}%
  \BibitemOpen
  \bibfield  {author} {\bibinfo {author} {\bibfnamefont {J.}~\bibnamefont
  {Bhattacharya}}, \bibinfo {author} {\bibfnamefont {S.}~\bibnamefont
  {Bhattacharyya}},\ and\ \bibinfo {author} {\bibfnamefont {S.}~\bibnamefont
  {Minwalla}},\ }\bibfield  {title} {\bibinfo {title} {{Dissipative Superfluid
  dynamics from gravity}},\ }\href {https://doi.org/10.1007/JHEP04(2011)125}
  {\bibfield  {journal} {\bibinfo  {journal} {JHEP}\ }\textbf {\bibinfo
  {volume} {04}}\bibfield  {number} {\bibinfo  {number} { (04)},\ \bibinfo
  {pages} {125}},\ }\Eprint {https://arxiv.org/abs/1101.3332} {arXiv:1101.3332
  [hep-th]} \BibitemShut {NoStop}%
\bibitem [{\citenamefont {Baggioli}\ and\ \citenamefont
  {Frangi}(2022)}]{Baggioli:2022aft}%
  \BibitemOpen
  \bibfield  {author} {\bibinfo {author} {\bibfnamefont {M.}~\bibnamefont
  {Baggioli}}\ and\ \bibinfo {author} {\bibfnamefont {G.}~\bibnamefont
  {Frangi}},\ }\bibfield  {title} {\bibinfo {title} {{Holographic
  Supersolids}},\ }\bibfield  {journal} {\bibinfo  {journal} {preprint}\ }\href
  {https://doi.org/10.1007/JHEP06(2022)152} {10.1007/JHEP06(2022)152} (\bibinfo
  {year} {2022}),\ \Eprint {https://arxiv.org/abs/2202.03745} {arXiv:2202.03745
  [hep-th]} \BibitemShut {NoStop}%
\bibitem [{\citenamefont {Baez}\ and\ \citenamefont
  {Huerta}(2011)}]{Baez:2010ya}%
  \BibitemOpen
  \bibfield  {author} {\bibinfo {author} {\bibfnamefont {J.~C.}\ \bibnamefont
  {Baez}}\ and\ \bibinfo {author} {\bibfnamefont {J.}~\bibnamefont {Huerta}},\
  }\bibfield  {title} {\bibinfo {title} {{An Invitation to Higher Gauge
  Theory}},\ }\href {https://doi.org/10.1007/s10714-010-1070-9} {\bibfield
  {journal} {\bibinfo  {journal} {Gen. Rel. Grav.}\ }\textbf {\bibinfo {volume}
  {43}},\ \bibinfo {pages} {2335} (\bibinfo {year} {2011})},\ \Eprint
  {https://arxiv.org/abs/1003.4485} {arXiv:1003.4485 [hep-th]} \BibitemShut
  {NoStop}%
\bibitem [{\citenamefont {Gaiotto}\ \emph {et~al.}(2015)\citenamefont
  {Gaiotto}, \citenamefont {Kapustin}, \citenamefont {Seiberg},\ and\
  \citenamefont {Willett}}]{Gaiotto:2014kfa}%
  \BibitemOpen
  \bibfield  {author} {\bibinfo {author} {\bibfnamefont {D.}~\bibnamefont
  {Gaiotto}}, \bibinfo {author} {\bibfnamefont {A.}~\bibnamefont {Kapustin}},
  \bibinfo {author} {\bibfnamefont {N.}~\bibnamefont {Seiberg}},\ and\ \bibinfo
  {author} {\bibfnamefont {B.}~\bibnamefont {Willett}},\ }\bibfield  {title}
  {\bibinfo {title} {{Generalized Global Symmetries}},\ }\href
  {https://doi.org/10.1007/JHEP02(2015)172} {\bibfield  {journal} {\bibinfo
  {journal} {JHEP}\ }\textbf {\bibinfo {volume} {02}}\bibfield  {number}
  {\bibinfo  {number} { (02)},\ \bibinfo {pages} {172}},\ }\Eprint
  {https://arxiv.org/abs/1412.5148} {arXiv:1412.5148 [hep-th]} \BibitemShut
  {NoStop}%
\bibitem [{nb-(2024{\natexlab{b}})}]{nb-hol}%
  \BibitemOpen
  \href@noop {} {\bibinfo {title} {See the {Mathematica} notebook}},\ \bibinfo
  {howpublished}
  {\url{https://github.com/ajainphysics/Mathematica-Notebooks/tree/main/\%5BarXiv\%3A2406.07334\%5D\%20Fractonic\%20Solids/FractonicSolids_Holography.nb}}
  (\bibinfo {year} {2024}{\natexlab{b}}),\ \bibinfo {note} {for details of the
  quasi-normal modes of a charged black brane in the holographic model for
  fractonic solids.}\BibitemShut {Stop}%
\bibitem [{\citenamefont {Grozdanov}\ and\ \citenamefont
  {Polonyi}(2015)}]{Grozdanov:2013dba}%
  \BibitemOpen
  \bibfield  {author} {\bibinfo {author} {\bibfnamefont {S.}~\bibnamefont
  {Grozdanov}}\ and\ \bibinfo {author} {\bibfnamefont {J.}~\bibnamefont
  {Polonyi}},\ }\bibfield  {title} {\bibinfo {title} {{Viscosity and
  dissipative hydrodynamics from effective field theory}},\ }\href
  {https://doi.org/10.1103/PhysRevD.91.105031} {\bibfield  {journal} {\bibinfo
  {journal} {Phys. Rev. D}\ }\textbf {\bibinfo {volume} {91}},\ \bibinfo
  {pages} {105031} (\bibinfo {year} {2015})},\ \Eprint
  {https://arxiv.org/abs/1305.3670} {arXiv:1305.3670 [hep-th]} \BibitemShut
  {NoStop}%
\bibitem [{\citenamefont {Harder}\ \emph {et~al.}(2015)\citenamefont {Harder},
  \citenamefont {Kovtun},\ and\ \citenamefont {Ritz}}]{Harder:2015nxa}%
  \BibitemOpen
  \bibfield  {author} {\bibinfo {author} {\bibfnamefont {M.}~\bibnamefont
  {Harder}}, \bibinfo {author} {\bibfnamefont {P.}~\bibnamefont {Kovtun}},\
  and\ \bibinfo {author} {\bibfnamefont {A.}~\bibnamefont {Ritz}},\ }\bibfield
  {title} {\bibinfo {title} {{On thermal fluctuations and the generating
  functional in relativistic hydrodynamics}},\ }\href
  {https://doi.org/10.1007/JHEP07(2015)025} {\bibfield  {journal} {\bibinfo
  {journal} {JHEP}\ }\textbf {\bibinfo {volume} {07}}\bibfield  {number}
  {\bibinfo  {number} { (07)},\ \bibinfo {pages} {025}},\ }\Eprint
  {https://arxiv.org/abs/1502.03076} {arXiv:1502.03076 [hep-th]} \BibitemShut
  {NoStop}%
\bibitem [{\citenamefont {Crossley}\ \emph {et~al.}(2017)\citenamefont
  {Crossley}, \citenamefont {Glorioso},\ and\ \citenamefont
  {Liu}}]{Crossley:2015evo}%
  \BibitemOpen
  \bibfield  {author} {\bibinfo {author} {\bibfnamefont {M.}~\bibnamefont
  {Crossley}}, \bibinfo {author} {\bibfnamefont {P.}~\bibnamefont {Glorioso}},\
  and\ \bibinfo {author} {\bibfnamefont {H.}~\bibnamefont {Liu}},\ }\bibfield
  {title} {\bibinfo {title} {{Effective field theory of dissipative fluids}},\
  }\href {https://doi.org/10.1007/JHEP09(2017)095} {\bibfield  {journal}
  {\bibinfo  {journal} {JHEP}\ }\textbf {\bibinfo {volume} {09}}\bibfield
  {number} {\bibinfo  {number} { (09)},\ \bibinfo {pages} {095}},\ }\Eprint
  {https://arxiv.org/abs/1511.03646} {arXiv:1511.03646 [hep-th]} \BibitemShut
  {NoStop}%
\bibitem [{\citenamefont {Haehl}\ \emph {et~al.}(2016)\citenamefont {Haehl},
  \citenamefont {Loganayagam},\ and\ \citenamefont
  {Rangamani}}]{Haehl:2015uoc}%
  \BibitemOpen
  \bibfield  {author} {\bibinfo {author} {\bibfnamefont {F.~M.}\ \bibnamefont
  {Haehl}}, \bibinfo {author} {\bibfnamefont {R.}~\bibnamefont {Loganayagam}},\
  and\ \bibinfo {author} {\bibfnamefont {M.}~\bibnamefont {Rangamani}},\
  }\bibfield  {title} {\bibinfo {title} {{Topological sigma models \&
  dissipative hydrodynamics}},\ }\href
  {https://doi.org/10.1007/JHEP04(2016)039} {\bibfield  {journal} {\bibinfo
  {journal} {JHEP}\ }\textbf {\bibinfo {volume} {04}}\bibfield  {number}
  {\bibinfo  {number} { (04)},\ \bibinfo {pages} {039}},\ }\Eprint
  {https://arxiv.org/abs/1511.07809} {arXiv:1511.07809 [hep-th]} \BibitemShut
  {NoStop}%
\bibitem [{\citenamefont {Haehl}\ \emph {et~al.}(2018)\citenamefont {Haehl},
  \citenamefont {Loganayagam},\ and\ \citenamefont
  {Rangamani}}]{Haehl:2018lcu}%
  \BibitemOpen
  \bibfield  {author} {\bibinfo {author} {\bibfnamefont {F.~M.}\ \bibnamefont
  {Haehl}}, \bibinfo {author} {\bibfnamefont {R.}~\bibnamefont {Loganayagam}},\
  and\ \bibinfo {author} {\bibfnamefont {M.}~\bibnamefont {Rangamani}},\
  }\bibfield  {title} {\bibinfo {title} {{Effective Action for Relativistic
  Hydrodynamics: Fluctuations, Dissipation, and Entropy Inflow}},\ }\href
  {https://doi.org/10.1007/JHEP10(2018)194} {\bibfield  {journal} {\bibinfo
  {journal} {JHEP}\ }\textbf {\bibinfo {volume} {10}}\bibfield  {number}
  {\bibinfo  {number} { (10)},\ \bibinfo {pages} {194}},\ }\Eprint
  {https://arxiv.org/abs/1803.11155} {arXiv:1803.11155 [hep-th]} \BibitemShut
  {NoStop}%
\bibitem [{\citenamefont {Jensen}\ \emph {et~al.}(2018)\citenamefont {Jensen},
  \citenamefont {Pinzani-Fokeeva},\ and\ \citenamefont
  {Yarom}}]{Jensen:2017kzi}%
  \BibitemOpen
  \bibfield  {author} {\bibinfo {author} {\bibfnamefont {K.}~\bibnamefont
  {Jensen}}, \bibinfo {author} {\bibfnamefont {N.}~\bibnamefont
  {Pinzani-Fokeeva}},\ and\ \bibinfo {author} {\bibfnamefont {A.}~\bibnamefont
  {Yarom}},\ }\bibfield  {title} {\bibinfo {title} {{Dissipative hydrodynamics
  in superspace}},\ }\href {https://doi.org/10.1007/JHEP09(2018)127} {\bibfield
   {journal} {\bibinfo  {journal} {JHEP}\ }\textbf {\bibinfo {volume}
  {09}}\bibfield  {number} {\bibinfo  {number} { (09)},\ \bibinfo {pages}
  {127}},\ }\Eprint {https://arxiv.org/abs/1701.07436} {arXiv:1701.07436
  [hep-th]} \BibitemShut {NoStop}%
\bibitem [{\citenamefont {Liu}\ and\ \citenamefont
  {Glorioso}(2018)}]{Glorioso:2018wxw}%
  \BibitemOpen
  \bibfield  {author} {\bibinfo {author} {\bibfnamefont {H.}~\bibnamefont
  {Liu}}\ and\ \bibinfo {author} {\bibfnamefont {P.}~\bibnamefont {Glorioso}},\
  }\bibfield  {title} {\bibinfo {title} {{Lectures on non-equilibrium effective
  field theories and fluctuating hydrodynamics}},\ }\href
  {https://doi.org/10.22323/1.305.0008} {\bibfield  {journal} {\bibinfo
  {journal} {PoS}\ }\textbf {\bibinfo {volume} {TASI2017}},\ \bibinfo {pages}
  {008} (\bibinfo {year} {2018})},\ \Eprint {https://arxiv.org/abs/1805.09331}
  {arXiv:1805.09331 [hep-th]} \BibitemShut {NoStop}%
\end{thebibliography}%

\end{document}